\documentclass[12pt,english,floatfix,nofootinbib,superscriptaddress,aps,prd,preprint]{revtex4}
\usepackage[utf8]{inputenc}
\usepackage{float}
\usepackage{array}
\usepackage{bbold}
\usepackage{lipsum}
\usepackage{dsfont}
\usepackage{graphicx}
\usepackage{amsmath,amsthm,amsfonts,amssymb}
\usepackage{graphicx}
\usepackage[english]{babel} 
\usepackage{color}
\usepackage{tensor}
\usepackage{esint}
\usepackage[dvips]{epsfig}
\usepackage[dvips]{graphicx}
\usepackage{float}
\usepackage{units}
\usepackage{textcomp}
\usepackage{mathrsfs}
\usepackage{amsmath}
\usepackage[makeroom]{cancel}
\usepackage{amssymb}
\usepackage{amsbsy}
\usepackage{amsfonts}
\usepackage{amssymb,mathrsfs,xcolor}
\usepackage{esint}
\usepackage{braket}
\usepackage{array}
\usepackage{graphicx}

\usepackage{wasysym}
\usepackage{multirow}
\usepackage{wrapfig}
\usepackage{subfig}

\usepackage{stmaryrd}
\usepackage{upgreek}

\makeatletter

\makeatletter\usepackage{babel}

\usepackage{hyperref}
\hypersetup{
    colorlinks,
    citecolor=blue,
    filecolor=green,
    linkcolor=purple,
    urlcolor=red,
}

\usepackage{slashed}

\newcommand{\ie}{\begin{equation}}
\newcommand{\fe}{\end{equation}}
\newcommand{\se}{\begin{eqnarray}}
\newcommand{\ff}{\end{eqnarray}}

\begin{document}

\title{Analysis of a regular black hole in Verlinde's gravity}

\author{A. A. Ara\'{u}jo Filho}
\email{dilto@fisica.ufc.br}

\affiliation{Departamento de Física Teórica and IFIC, Centro Mixto Universidad de Valencia--CSIC. Universidad
de Valencia, Burjassot--46100, Valencia, Spain}

\affiliation{Departamento de Física, Universidade Federal da Paraíba, Caixa Postal 5008, 58051-970, João Pessoa, Paraíba,  Brazil.}


\date{\today}

\begin{abstract}

This work focuses on the examination of a regular black hole within Verlinde's emergent gravity, specifically investigating the Hayward--like (modified) solution. The study reveals the existence of three horizons under certain conditions, i.e., an event horizon and two Couchy horizons. Our results indicate regions which phase transitions occur based on the analysis of heat capacity and \textit{Hawking} temperature.  To compute the latter quantity, we utilize three distinct methods: the surface gravity approach, \textit{Hawking} radiation, and the application of the first law of thermodynamics. In the case of the latter approach, it is imperative to introduce a correction to ensure the preservation of the \textit{Bekenstein--Hawking} area law. Geodesic trajectories and critical orbits (photon spheres) are calculated, highlighting the presence of three light rings. Additionally, we investigate the black hole shadows. Furthermore, the \textit{quasinormal} modes are explored using third-- and sixth--order WKB approximations. In particular, we observe stable and unstable oscillations for certain frequencies. Finally, in order to comprehend the phenomena of time--dependent scattering in this scenario, we provide an investigation of the time--domain solution.

\end{abstract}

\maketitle


\section{Introduction}

General Relativity is known to be incomplete in both the classical and quantum realms, particularly due to the presence of singularities. Classical Einstein's theory of gravity faces the challenge of inevitable singularities in solutions such as Schwarzschild, Reisner--Nordström, and Kerr metrics, which exhibit such a feature within their interiors. It is widely accepted that modifications to General Relativity are necessary in regions where spacetime curvature reaches high values.

Such a modification is necessary not only to address the existence of singularities but also to achieve a theory that is ultraviolet (UV) complete. Several proposals have been put forward to accomplish this modification. It has been demonstrated that the inclusion of higher--order curvature terms, as well as terms involving higher derivatives, can enhance the UV properties of Einstein's gravity \cite{1,2,3,4}.

However, these modified theories often have the presence of non--physical degrees of freedom, known as ghosts. In recent years, a novel UV--complete modification of General Relativity has been proposed, addressing this issue effectively \cite{5,6,7}. Referred to as ghost--free gravity \cite{5,6,7,8,9,10,12,13,14}, this theory incorporates an infinite number of derivatives and exhibits non--local characteristics \cite{10,12,13}. Interestingly, a similar theory arises naturally within the framework of noncommutative geometry deformation of Einstein's gravity \cite{15,16}, as discussed in a comprehensive review \cite{17} and its references. The application of ghost--free gravity to the study of singularities in cosmology and black holes has been explored in \cite{18,19,20,21,22,23,24,25}. 

In the absence of a specific theory, it is informative to consider the potential modifications that could arise when gravity is UV complete. Exploring such possibilities can be valuable, provided that certain ``natural" assumptions regarding the properties of a comprehensive theory are imposed. In this regard, we focus on regular(non--singular) models of black holes. In other words, the goal is to explore black hole metrics that are devoid of curvature singularities. The pioneering work of Bardeen \cite{26} introduced the concept of a non--singular black hole, where the singularity is replaced by a charged matter core resulting from the collapse of charged matter. Subsequently, various models of non--singular black holes, including neutral, charged, and rotating configurations, have been proposed and discussed \cite{27,28,29,30,31,32,33,34,35,36,37,38,39,frolov2016notes,neves2014regular,maluf2018thermodynamics,neves2017deforming,neves2017bouncing,neves2020accretion,maluf2019bardeen,40}.

On the other hand, a comprehensive understanding of gravitational waves and their properties is crucial for the investigation of various physical phenomena, ranging from cosmological events in the early universe to astrophysical processes such as the evolution of stellar oscillations \cite{unno1979nonradial, kjeldsen1994amplitudes, dziembowski1992effects} and binary systems \cite{pretorius2005evolution, hurley2002evolution, yakut2005evolution, heuvel2011compact}. They exhibit a diverse range of intensities and characteristic modes, with their spectral characteristics being influenced by the underlying phenomena that give rise to them \cite{riles2017recent}. When a black hole is formed through the gravitational collapse of matter, it undergoes a perturbed state and emits radiation that encompasses a set of distinct frequencies unrelated to the collapse itself \cite{konoplya2011quasinormal}. These perturbations, characterized by specific frequencies, are commonly known as \textit{quasinormal} modes \cite{heidari2023gravitational, kokkotas1999quasi}.

The study of \textit{quasinormal} modes of black holes has been extensively conducted in the literature, employing the weak field approximation. This approach has been applied not only in the framework of GR \cite{rincon2020greybody, santos2016quasinormal, oliveira2019quasinormal, berti2009quasinormal, horowitz2000quasinormal, nollert1999quasinormal, ferrari1984new, kokkotas1999quasi, london2014modeling, maggiore2008physical, flachi2013quasinormal, ovgun2018quasinormal, blazquez2018scalar, roy2020revisiting, konoplya2011quasinormal}, but also in alternative gravity theories including Ricci--based theories \cite{kim2018quasi, lee2020quasi, jawad2020quasinormal}, Lorentz violation \cite{maluf2013matter, maluf2014einstein}, and other related fields \cite{JCAP1, JCAP2, JCAP3, jcap4, jcap5}.

Remarkable advancements have been achieved in the advancement of gravitational wave detectors, facilitating the identification of them emitted by various physical phenomena \cite{abbott2016ligo, abbott2017gravitational, abbott2017gw170817, abbott2017multi}. Crucial contributions to these detections have been made by ground--based interferometers such as VIRGO, LIGO, TAMA--300, and EO--600 \cite{fafone2015advanced, abramovici1992ligo, coccia1995gravitational, luck1997geo600}. Over time, these detectors have significantly enhanced their precision, approaching the realm of genuine sensitivity \cite{evans2014gravitational}. The insights garnered from these ones have provided valuable knowledge regarding the nature of astrophysical entities, encompassing boson and neutron stars.

The detection of gravitational waves has profound implications for the field of black hole physics. Through the observation of emitted gravitational radiation, the existence of perturbed black holes can be directly confirmed \cite{thorne2000probing}. Pioneering investigations into black hole perturbations were carried out by Regge and Wheeler, who explored the stability of Schwarzschild black holes \cite{regge1957stability}, followed by Zerilli's seminal contributions to the study of perturbations \cite{zerilli1970effective, zerilli1974perturbation}.

In recent years, there has been significant interest in the study of gravitational solutions involving scalar fields, driven by their intriguing and distinctive characteristics. These investigations have led to a wide range of astrophysical applications. One notable area of exploration is the behavior of black holes with nontrivial scalar fields, which seems to contradict the well--established ``no--hair theorem" \cite{herdeiro2015asymptotically}. Furthermore, the existence of long--lived scalar field patterns \cite{ayon2016analytic}, the formation of boson stars \cite{colpi1986boson, palenzuela2017gravitational, cunha2017lensing}, and the examination of exotic astrophysical scenarios such as gravastars \cite{visser2004stable, pani2009gravitational, chirenti2016did} have all emerged from this line of research. Additionally, considering Klein--Gordon scalar fields on curved backgrounds unveils a wide range of phenomena, including black hole bombs \cite{cardoso2004black, sanchis2016explosion, hod2016charged} and superradiance \cite{brito2015black}. 

In particular, according to Verlinde's theory \cite{verlinde2011origin}, dark matter can be understood as an emergent consequence of gravity arising from the distribution of baryonic matter. Verlinde postulates the existence of an additional gravitational effect caused by the volume law contribution to entropy associated with positive dark energy. This implies that the distribution of baryonic matter reduces the overall entropy of the universe, resulting in an elastic response of the underlying microscopic system. Consequently, this elastic response generates an extra gravitational effect, commonly referred to as the dark matter effect, which is inherent to gravity itself. From an observational perspective, Verlinde's theory suggests that the relationship between the distribution of baryonic matter and apparent dark matter can account for the observed flat rotation curves observed in galaxies \cite{verlinde2017emergent}.

In this paper, we undertake a comprehensive examination of a regular black hole within the framework of Verlinde's emergent gravity, with a specific focus on investigating the modified Hayward--like solution. Our rigorous analysis reveals intriguing findings, notably the existence of three distinct horizons under specific conditions: an event horizon and two Couchy horizons. Furthermore, our research extends to the scrutiny of critical regions where phase transitions occur, as evidenced through a meticulous study of heat capacity and the Hawking temperature. To compute the Hawking temperature, we employ three distinct methods: the surface gravity approach, \textit{Hawking} radiation, and the application of the first law of thermodynamics. In doing so, we introduce a correction to uphold the integrity of the \textit{Bekenstein--Hawking} area law. Expanding our inquiry, we explore the calculation of geodesic trajectories and critical orbits, which intriguingly unveil the presence of three distinct light rings. Additionally, our investigation encompasses the analysis of black hole shadows. Going further, we examine quasinormal modes using third and sixth--order WKB approximations, revealing both stable and unstable oscillations for specific frequency ranges. Finally, to comprehend the intricacies of time--dependent scattering phenomena in this scenario, we provide a comprehensive examination of the time--domain solution.

\section{Regular black hole in Verlinde's gravity}

The historical antecedents of the thermodynamic elucidation of gravity can be traced back to the pioneering research conducted in the mid--1970s by Bekenstein \cite{bekenstein1973black,bekenstein1974generalized,bekenstein2020black} and Hawking \cite{bardeen1973four} on the thermodynamics of black holes. Their investigations revealed a profound correlation between the gravitational force and the principles governing thermodynamics, which encompasses the study of heat behavior.

Significantly, in 1995, Jacobson advanced our understanding by establishing that the Einstein field equations, the fundamental mathematical framework that characterizes relativistic gravitation, can be derived through the amalgamation of overarching thermodynamic principles with the foundational concept of the equivalence principle \cite{jacobson1995thermodynamics}. This groundbreaking revelation marked a pivotal milestone in the synthesis of gravitational theory and thermodynamics. Following this pivotal research, Padmanabhan continued to investigate the connection between gravity and the concept of entropy \cite{padmanabhan2010thermodynamical}.

In 2009, Verlinde introduced a groundbreaking conceptual model positing gravity as an entropic force. This theory, akin to Jacobson's findings, asserts that gravity results from the information associated with the positions of material bodies \cite{017}. Verlinde's model amalgamates the thermodynamic perspective on gravity with Hooft's holographic principle \cite{hooft1999quantum,hooft2001holographic}, fundamentally contending that gravity is not a fundamental interaction but rather an emergent phenomenon stemming from the statistical behavior of microscopic degrees of freedom encoded on a holographic screen.

Subsequently, in 2011, Verlinde expounded upon his theories, including an explanation for the genesis of dark matter \cite{071,07}. This work garnered considerable media attention and prompted immediate research initiatives across cosmology \cite{010,011}, cosmological acceleration \cite{013,014}, cosmological inflation \cite{015}, and loop quantum gravity \cite{016}. Furthermore, specific microscopic models have been advanced, demonstrating the emergence of entropic gravity on larger scales, with entropic gravity being shown to result from quantum entanglement of local Rindler horizons \cite{017,018}. Furthermore, recent literature has extensively explored the theoretical implications \cite{wang2018surfaces,liu2017gravitational,buchel2017verlinde} as well as the observational consequences \cite{pardo2020testing,tamosiunas2019testing,brouwer2021weak} of Verlinde's emergent gravity.

In our present work, our primary objective is to utilize the connection between baryonic matter and apparent dark matter to construct black hole solutions within the framework of this theory, known as VEG (Verlinde's emergent gravity). By doing so, we aim to explore the additional effects stemming from the presence of apparent dark matter on the geometry of spacetime in the next sections.

Verlinde argues that under the assumption of spherical symmetry, there exists a relationship between the quantities of apparent dark and baryonic matters \cite{jusufi2023regular}, i.e., $M_{D}(r)$ and $M_{B}(r)$, respectively. This relationship can be expressed through the following expression:
\ie
\int^{r}_{0} \frac{M^{2}_{D}(\tilde{r})}{\tilde{r}^{2}}\mathrm{d}\tilde{r} = \frac{a_{0}M_{B}(r)r}{6},
\fe
where $a_{0}$ is a constant. In the simplest scenario, we consider spherically symmetric black hole solutions described by the line element
\ie
\mathrm{d}s^{2} = g_{\mu\nu}\mathrm{d}x^{\mu}\mathrm{d}x^{\nu}= g_{00}\mathrm{d}t^{2} + g_{11}\mathrm{d}r^{2} + g_{22}\mathrm{d}\theta^{2} + g_{33}\mathrm{d}\phi^{2},
\label{thefundamentalmetric}
\fe
with $f(r)=-g_{00}=g^{-1}_{11}= 1 - 2m(r)/r$, $g_{22}=r^{2}$, $g_{33}=r^{2}\sin^{2}\theta$, and $m(r)$, being 
\ie
m(r) = 4\pi \int^{r}_{0} \left[ \rho_{B}(\tilde{r}) + \rho_{D}(\tilde{r}) \right]\tilde{r}^{2} \mathrm{d}\tilde{r}.
\fe
Based on the recent results in Ref. \cite{jusufi2023regular}, we shall obtain the a regular black hole solution. In the next sections, we shall see the dark matter effects on the geometry of the spacetime. Also, we analyze the signatures brought about for such a spacetime, i.e., the event horizon, the thermodynamic, the geodesics, the shadows, and the \textit{quasinormal} modes.


\section{Thermodynamics}

\begin{figure}
    \centering
    \includegraphics[scale=0.4]{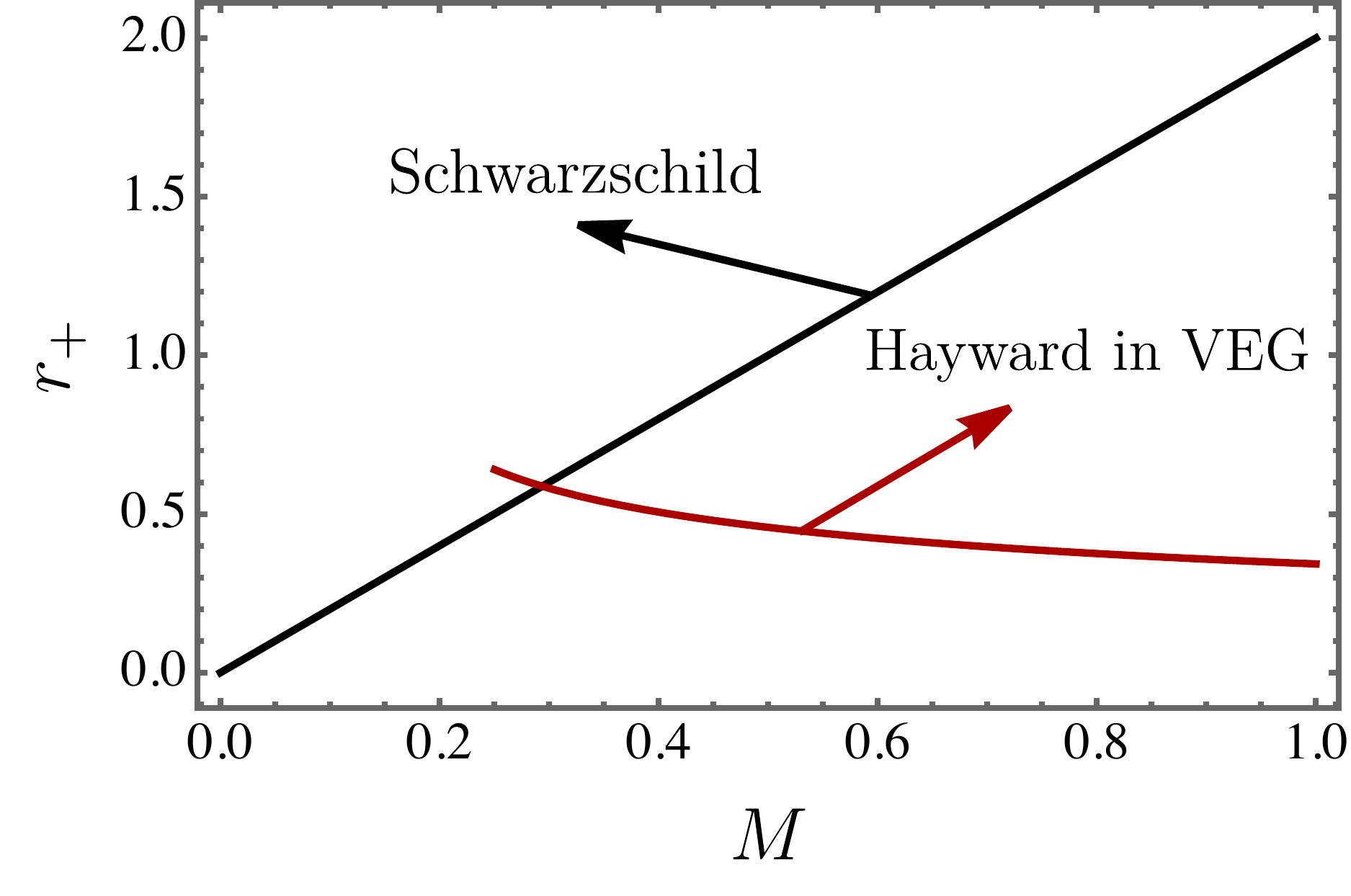}
    \caption{The comparison between the Schwarzschild and modified Hayward horizons.}
    \label{eventhorizon}
\end{figure}

In this section, we regard the Hayward mass function, which reads \cite{hayward2006formation}
\ie
M_{B}(r) = \frac{M r^{3}}{r^{3}+a^{3}},
\fe
where $a$ is a constant with dimension of length. Here, one interesting question gives rise to: what would be the mass function for considering a more general case with a charge $Q$? This answer was addressed by Frolov in Ref. \cite{frolov2016notes}. Although such an aspect is worthy to be investigated, it lies beyond the scope of the current work. Nevertheless, these and other ideas will be addressed in forthcoming manuscripts. Moreover, solving for the apparent dark matter, we obtain \cite{jusufi2023regular}
\ie
f(r) = 1 -\frac{2 M r^2}{a^3+r^3} - \frac{2 \sqrt{a_{0} M} \sqrt{r^3 \left(4 a^3+r^3\right)}}{a^3+r^3}. 
\label{frrr}
\fe
In spite of having six roots, above expression leads to one physical horizon only (for $a>0.31977$ and $a_{0}=1$):
\ie
\begin{split}
r_{+} &= -\frac{1}{2} \sqrt{\frac{16 \sqrt[3]{2} a^6}{\frac{\chi }{12 \sqrt[3]{2} M^2}+3 \chi }+\frac{2 a^3}{3 M}+\frac{\eta ^2}{16 M^4}} -\frac{\eta }{8 M^2} + \mathcal{O}(M^{5}) +  \mathcal{O}(M^{6})\\
& + \frac{1}{2} \sqrt{-\frac{16 \sqrt[3]{2} a^6}{3 \chi }+\frac{4 a^3}{3 M}-\frac{-\frac{2 a^3 \eta }{M^3}-\frac{\eta ^3}{8 M^6}}{4 \sqrt{\frac{16 \sqrt[3]{2} a^6}{3 \chi }+\frac{2 a^3}{3 M}+\frac{\eta ^2}{16 M^4}+\frac{\chi }{12 \sqrt[3]{2} M^2}}}+\frac{\eta ^2}{8 M^4}-\frac{\chi }{12 \sqrt[3]{2} M^2}}
\end{split}
\fe
where, 
\ie
\eta =a^3-8 a^3 a_0 M,\nonumber
\fe
\ie
\chi =\sqrt[3]{1024 a^9 M^3+108 a^6 \eta ^2+\gamma },\nonumber
\fe
and 
\ie
\gamma =\sqrt{\left(1024 a^9 M^3+108 a^6 \eta ^2\right)^2-1048576 a^{18} M^6}.\nonumber
\fe
It is important to mention that, besides we have written higher--order terms in a compact form, namely, $\mathcal{O}(M^{5})$, and $ \mathcal{O}(M^{6})$, they will certainly be taken into account in our calculations.
In addition, to have real positive defined values, the mass has to possess the following condition $M>0.13705$ (if $a_{0}=a=1$). The theory under consideration has a remarkable feature worth mentioning: considering values of parameter $a$ between $0<a\leqslant 0.31977$ (for $a_{0}=1$),  Eq. (\ref{frrr}) possesses three different solutions, namely, $r_{+}$, $r_{\cdot}$, and $r_{-}$. In other words, it might be interpreted as an event horizon (the outermost)  and two killing horizons (the innermost) respectively. In order to provide a better comprehension to the reader, we present the Fig. \ref{eventhorizon}. Here, we plot the $r_{+}$ and the Schwarzschild horizon as a function of mass $M$.
For the sake of bring more quantitative more information about event horizon, we present the Tables \ref{horizongreaterthana1} and \ref{horizonintermediates} for both cases: $a > 0.31977$ and $0<a\leqslant 0.31977$, respectively. In the first scenario, it becomes evident that as the mass parameter $M$ increases, the size of the event horizon decreases, all while keeping parameter $a$ constant. Conversely, when we vary the angular momentum parameter $a$ (while keeping $M$ constant), the event horizon exhibits the opposite behavior.

Turning our attention to the second scenario, as we can observe in Table \ref{horizonintermediates}, an increase in mass leads to a decrease in the values of $r_{-}$ and $r_{\cdot}$ (for a constant $a$). Conversely, when we vary $M$ (with $a$ held constant), $r_{+}$ increases. On the other hand, when parameter $a$ changes (while maintaining $M$ constant), both $r_{-}$ and $r_{\cdot}$ increase, and likewise, $r_{+}$ increases as $M$ varies.

Furthermore, in the next subsections, we shall devote our attention on thermodynamic properties of the system. Particularly, we shall calculate the temperature, the entropy, and the heat capacity.

\begin{table}[!h]
\begin{center}
\begin{tabular}{c c c} 
 \hline\hline
 $M$ & $a$ & $r_{c_{+}}$  \\ [0.2ex] 
 \hline 
 1.0  & 1.0 & 0.342 \\ 

 2.0  & 1.0 & 0.259  \\
 
 3.0  & 1.0 & 0.221  \\
 
 4.0  & 1.0 & 0.197  \\
 
 5.0  & 1.0 & 0.180  \\
 
 6.0  & 1.0 & 0.167  \\
 
 7.0  & 1.0 & 0.157  \\
 
 8.0  & 1.0 & 0.149  \\
 
 9.0  & 1.0 & 0.142  \\
 
 10.0  & 1.0 & 0.136  \\
 [0.2ex] 
 \hline \hline
  \hline
  $M$ & $a$ & $r_{c_{+}}$  \\ [0.2ex] 
 \hline 
 1.0  & 1.0 & 0.342  \\ 

 1.0  & 2.0 & 0.744  \\
 
 1.0  & 3.0 & 1.150   \\
 
 1.0  & 4.0 & 1.560  \\
 
 1.0 & 5.0 & 1.970  \\
 
 1.0  & 6.0 & 2.380  \\
 
 1.0  & 7.0 & 2.800  \\
 
 1.0  & 8.0 & 3.210  \\
 
 1.0  & 9.0 & 3.620  \\
 
 1.0  & 10.0 & 4.030  \\
 [0.2ex] \hline \hline \hline
\end{tabular}
\caption{\label{horizongreaterthana1}The values of the horizon $r_{+}$ (when $a >  0.31977$) for different values of mass $M$ and parameter $a$.}
\end{center}
\end{table}

\begin{table}[!h]
\begin{center}
\begin{tabular}{c c c c c} 
 \hline\hline
 $M$ & $a$ & $r_{-}$ & $r_{\cdot}$ & $r_{+}$   \\ [0.2ex] 
 \hline 
 1.0  & 0.1 & 0.01853330 & 0.0286133 & 0.6628840 \\ 

 2.0  & 0.1 & 0.01344790 & 0.0192766 & 1.0432200  \\
 
 3.0  & 0.1 & 0.01113490 & 0.0153971 & 1.3430700 \\
 
 4.0  & 0.1 & 0.00973365 & 0.0131529 & 1.5993000  \\
 
 5.0  & 0.1 & 0.00876641 & 0.0116505 & 1.8269000  \\
 
 6.0  & 0.1 & 0.00804613 & 0.0105565 & 2.0338100 \\
 
 7.0  & 0.1 & 0.00748237 & 0.0097153 & 2.2248500  \\
 
 8.0  & 0.1 & 0.00702526 & 0.0090430 & 2.4032200  \\
 
 9.0  & 0.1 & 0.00664474 & 0.0084901 & 2.5711500  \\
 
 10.0  & 0.1 & 0.00632142 & 0.0080253 & 2.7302900  \\
 [0.2ex] 
 \hline \hline
  \hline
  $M$ & $a$ & $r_{-}$ & $r_{\cdot}$ & $r_{+}$  \\ [0.2ex] 
 \hline 
 1.0  & 0.10 & 0.0185333  & 0.0286133 & 0.662884   \\ 

 1.0  & 0.11 & 0.0211201  &  0.0337433  & 0.661617  \\
 
 1.0  & 0.12 & 0.0237810  & 0.0393082 & 0.660088   \\
 
 1.0  & 0.13 & 0.0265097 &  0.0453262  &  0.658268   \\
 
 1.0 & 0.14 &  0.0293005  &  0.0518189  & 0.656124  \\
 
 1.0  & 0.15 & 0.0321487 &  0.0588115  &  0.653625   \\
 
 1.0  & 0.16 &  0.0350500  & 0.0663335  & 0.650731  \\
 
 1.0  & 0.17 & 0.0380005  & 0.0744194  & 0.647403  \\
 
 1.0  & 0.18 & 0.0409968 &  0.0831098  & 0.643594   \\
 
 1.0  & 0.19 &  0.0440361  & 0.0924526  & 0.639251  \\

 1.0  & 0.20 & 0.0471154  &  0.1025050  & 0.634311  \\
 [0.2ex] \hline \hline \hline
\end{tabular}
\caption{\label{horizonintermediates} The values of the horizons $r_{+}$, $r_{\cdot}$, and $r_{-}$ (when $0<a\leqslant 0.31977$) for different values of mass $M$ and parameter $a$.}
\end{center}
\end{table}


\subsection{The Hawking temperature}

Regarding temperature, we present three distinct methodologies for its determination: via surface gravity denoted as $\kappa$, through the tunneling method, and employing the first law of thermodynamics in a manner that parallels what one encounters in Ref. \cite{maluf2018thermodynamics}. Notably, the first law of black hole thermodynamics necessitates a refinement when applied to regular black holes \cite{ma2014corrected} in order to get a proper second law of thermodynamics and consequently the area law for entropy. Here, it is worth mentioning that besides the usual equation used for the first law of thermodynamic gives the correct result for the \textit{Hawking} temperature, it does not provide the correct area law for entropy as argued in Ref. \cite{ma2014corrected}.


\subsubsection{Via surface gravity}

The metric defined in Eq. (\ref{thefundamentalmetric}) possesses a timelike Killing vector, denoted as $\xi = \partial/\partial t$. Consequently, this metric gives rise to a conserved quantity that is inherently linked to $\xi$. We can indeed construct such a conserved quantity by leveraging this Killing vector in the following manner:
\ie
\nabla^\nu(\xi^\mu\xi_\mu) = -2\kappa\xi^\nu.
\fe
In this context, \(\nabla_\nu\) represents the covariant derivative, and \(\kappa\) remains unchanging along the orbits defined by \(\xi\). Specifically, this implies that the Lie derivative of \(\kappa\) with respect to \(\xi\) is identically zero:
\ie
\mathcal{L}_\xi\kappa = 0.
\fe
Notably, $\kappa$ remains constant across the horizon and it is referred to as the surface gravity. In the coordinate basis, the timelike Killing vector components are expressed as $\xi^{\mu} = (1, 0, 0, 0)$ . The surface gravity for the metric ansatz of Eq. (\ref{thefundamentalmetric}) is represented as follows: 
\ie 
\kappa = {\left.\frac{f^{\prime}(r)}{2} \right|_{r = {r_{+}}}}.
\fe 
Hawking's pioneering work, as presented in Ref. \cite{hawking1975particle}, revealed that black holes emit radiation. This radiation is associated with a temperature known as the Hawking temperature, which is expressed as $T = \kappa/2\pi$. To our case, it reads
\ie 
T = \frac{r_{+} \left(r_{+}^3-2 a^3\right) \left(M \sqrt{4 a^3 r_{+}^3+r_{+}^6}+3 a^3 r_{+} \sqrt{a_0 M}\right)}{\pi  \left(a^3+r_{+}^3\right)^2 \sqrt{4 a^3 r_{+}^3+r_{+}^6}}. \label{tem1}
\fe

\begin{figure}
    \centering
    \includegraphics[scale=0.3]{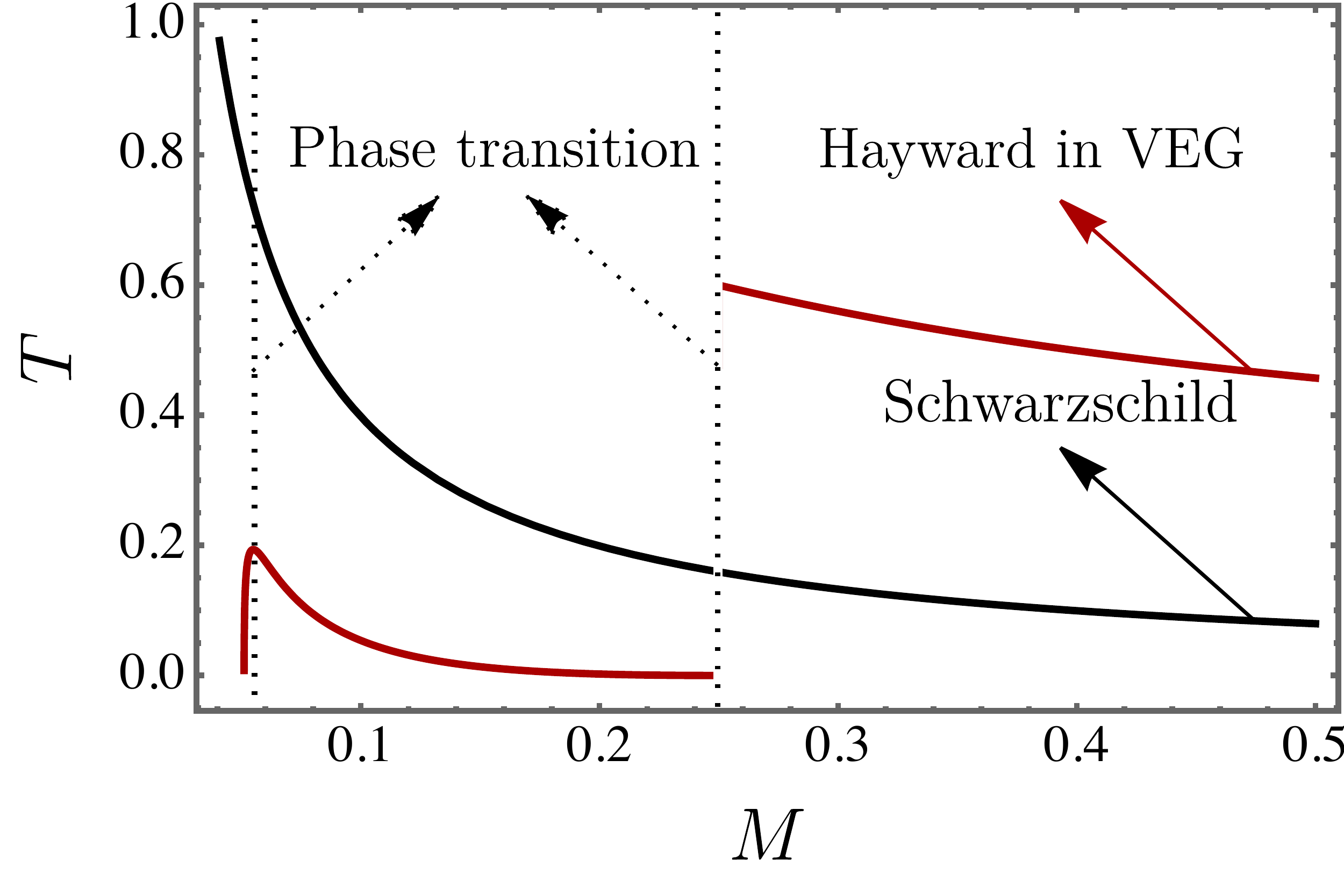}
    \caption{The \textit{Hawking} temperature in comparison with the Schwarzschild case. Two regions of phase transitions are present.}
    \label{temperature}
\end{figure}
In Fig. \ref{temperature}, we show the behavior of the \textit{Hawking} temperature as a function of mass. Notably, our results seem to indicate that when $M\approx 0.054$, and $M \approx 0.249$ there occur phase transitions. This feature should be a direct implication of the dark matter effects on spherically symmetric black hole under consideration.

\subsubsection{Via Hawking radiation}

As argued in the previous subsections, there also exist another method for addressing the \textit{Hawking} temperature: by considering the tunneling effect. Such an effect is also known in the literature as the \textit{Hawking} radiation.

In other words, the quantum tunneling phenomenon permits particles within the black hole to traverse the event horizon. The tunneling probability of this process can be computed, as demonstrated in Refs. \cite{angheben2005hawking,kerner2006tunnelling,kerner2008fermions}. In this particular approach, our focus is on radial trajectories. As a result, the metric under examination can be effectively treated as two--dimensional in the vicinity of the horizon
\ie
\mathrm{d}s^{2} = -f(r) \mathrm{d}t^{2} + \frac{\mathrm{d}r^{2}}{f(r)}.
\fe
Therefore, the issue is fully resolved within the $t-r$ plane. We can describe the Klein--Gordon equation in curved spacetime for a scalar field $\varphi$ with mass $m_{\varphi}$ as follows:
\ie
 \hbar^{2} g^{\mu\nu} \nabla_\mu \nabla_\nu \varphi - m^2_{\varphi} \varphi = 0,
\fe
which, it reads
\ie
-\partial_t^2 \varphi + f(r)^2 \partial_r^2 \varphi + \frac{1}{2} \partial_{r} f(r)^2 \partial_r \varphi - \frac{m^{2}_{\varphi}}{\hbar}f(r)\varphi = 0.
\fe
When employing the Wentzel--Kramers--Brillouin (WKB) method, the solution for the above equation can be written as:
\ie
\varphi(t,r) = e^{-\frac{i}{\hbar} I(t,r)}.
\fe
After that, we can write the Hamilton--Jacobi equation below
\ie
\left(\partial_t I\right)^2 - f(r)^2 \left(\partial_r I\right)^2 - m_{\varphi}^2 f(r) = 0,
\fe
where $I(t,r) = -Et + W(r)$ with $E$ being the radiation energy and $W(r)$ as
\ie
W_{\pm}(r) = \pm \int \mathrm{d}r \frac{1}{f(r)} \sqrt{E^{2} - m_{\varphi}^{2} f(r)} \label{dablui}.
\fe
Here the ``$+$'' and ``$-$'' represent the outgoing and the ingoing solutions respectively. In the classical context, $W_{+}(r)$ is typically prohibited as it represents solutions that cross the event horizon, moving away from $r_{+}$. Therefore, in order to explore Hawking radiation beyond the event horizon, our attention will be directed towards $W_{+}(r)$. Given the approximation for the function $f(r)$ in the vicinity of the event horizon at around $r_{+}$, we have
\ie
f(r) = f(r_{+}) + f^{\prime}(r_{+})(r-r_{+}) + ..., 
\fe
in a such way that Eq. (\ref{dablui}) is reduced to
\ie
W_{+}(r) = \frac{2i\pi E}{f^{\prime}(r_{+})}.
\fe
Hence, in the case of a particle, the probability of tunneling across the event horizon is determined by the imaginary component of the expression of $I(t,r)$. In other words,
\ie
\Gamma \simeq e^{-2\text{Im}I} = e^{-\frac{4\pi E}{f^{\prime}(r_{+})}}.
\fe
Drawing a comparison between the above expression and the Boltzmann factor $e^{-E/T}$, we can express the \textit{Hawking} temperature obtained through the tunneling method as:
\ie
\Tilde{T} = \frac{E}{2 \text{Im}I} = \frac{f^{\prime}(r_{+})}{4\pi}. \label{tem2}
\fe
Remarkably, $\Tilde{T}$ is fundamentally equal to the \textit{Hawking} temperature derived via the usage of the surface gravity $T$.



\subsection{Entropy}

Notice that a precise expression for $M$ is obtained when we consider $f(r_{+})=0$. Thereby, it can be written as
\ie
M_{\pm} = \frac{a^3 r_{+}^2+a_0 \left(4 a^3 r_{+}^3+r_{+}^6\right) \pm \sqrt{a_0 r_{+}^5 \left(4 a^3+r_{+}^3\right) \left(2 \left(a^3+r_{+}^3\right)+a_0 r_{+} \left(4 a^3+r_{+}^3\right)\right)}+r_{+}^5}{2 r_{+}^4},
\label{massequation}
\fe
where, notably, $M_{\pm}$ possesses two distinct solutions. This features rises fundamentally due to the square root presented in Eq. (\ref{frrr}). However, after performing some careful substitutions into Eq. (\ref{massequation}), concerning different values for $r_{+}$, and $a$ (keeping $a_{0}=1$), just one of them turns out to be a physical solution, i.e., $M_{-}$. Furthermore, this alignment with our interpretation is also supported by the fact that the deviation from the Schwarzschild solution is the smallest.

In addition, via the first law of thermodynamics, the \textit{Hawking} temperature is defined as
\ie
\Tilde{T}_{f}  = \frac{\mathrm{d}M_{-}}{\mathrm{d}S}= \frac{1}{2\pi r_{+}} \frac{\mathrm{d}M_{-}}{\mathrm{d}r_{+}} =  \frac{\left(r_{+}^3-2 a^3\right) \left(-2 a^3 (4 r_{+}+3) r_{+}^3+\alpha -2 r_{+}^7-3 r_{+}^6+2 \alpha  r_{+}\right)}{4 \pi \alpha  r_{+}^4},
\fe
with 
\ie
\alpha = \sqrt{r_{+}^5 \left(4 a^3+r_{+}^3\right) \left(a^3 (4 r_{+}+2)+r_{+}^3 (r_{+}+2)\right)}.
\nonumber
\fe
This temperature clearly disagrees with those ones encountered in Eqs. (\ref{tem1}) and (\ref{tem2}).
As pointed out previously, in accordance with the methodology outlined in Ref. \cite{ma2014corrected} and assuming the validity of the area law, we apply the corrected first law of thermodynamics that establishes the temperature for the entire category of regular black roles.

The corrected temperature is deduced from the following first law:
\ie
\Upsilon(r_{+},a) \mathrm{d}M = \overset{\nsim}{T}_{f}\, \mathrm{d}S,
\fe
where $\overset{\nsim}{T}_{f}$ is the corrected version of the \textit{Hawking} temperature calculated via first law of thermodynamics, and $S$ is the entropy. 
It is worth noting that the function $\Upsilon(r_{+}, a)$, which relies on the terms of the mass function, not only determines the first law for regular black holes in the specific case discussed here; besides, it plays a pivotal role in establishing the first law for various other classes of regular black holes \cite{maluf2018thermodynamics}.

As pointed out in Ref. \cite{ma2014corrected}, the general formula to $\Upsilon(r_{+},a)$ is given by
\ie
\Upsilon(r_{+},a) = 1 + 4\pi \int^{\infty}_{r_{+}} r^{2} \frac{\partial T^{0}_{0}}{\partial  M_{-}} \mathrm{d}r.
\fe
The notation $T^{0}_{0}$ pertains to the stress--energy component corresponding to energy density. Explicitly, $T^{0}_{0} = 1/8\pi G^{0}_{0}$ is given by
\ie
\begin{split}
T^{0}_{0} = &\frac{1}{8\pi\left(a^3+r^3\right)^3 \sqrt{4 a^3 r^3+r^6}} \left\{ 20 a^9 r \sqrt{a_0 M}-4 a_0 M r \left(10 a^6+2 a^3 r^3+r^6\right) \sqrt{4 a^3 r^3+r^6} \right. \\
&\left.  +2 r^9 \sqrt{a_0 M} (r-2 M) + a^6 \left(6 M \sqrt{4 a^3 r^3+r^6}+24 r^4 \sqrt{a_0 M}-88 M r^3 \sqrt{a_0 M}\right) \right. \\
& \left. a^3\left(-12 M^2 r^2 \sqrt{4 a^3 r^3+r^6}+6 M r^3 \sqrt{4 a^3 r^3+r^6}+6 r^7 \sqrt{a_0 M}-20 M r^6 \sqrt{a_0 M}\right.\right\}.
\end{split}
\fe
Naturally, in instances where the energy density remains unaffected by variations in mass, the correction term becomes negligible. In particular, to our case, we can explicitly express this correction term as follows:
\ie
\begin{split}
& \Upsilon(r_{+},a) = 1 +  4\pi \int^{\infty}_{r_{+}} r^{2} \left\{ \left[-8 \sqrt{2} a^9 r^3 (33 r+7)-r^7 \left(-3 \sqrt{2} \alpha +3 \sqrt{2} r^6+2 \sqrt{2} r^5+4 r^2 \psi  \omega \right) \right.\right.\\ 
& \left.\left. -a^6 r \left(-66 \sqrt{2} \alpha +126 \sqrt{2} r^6+69 \sqrt{2} r^5+88 r^2 \psi  \omega +6 r \psi  \omega \right) \right.\right. \\
& \left.\left. +a^3 \left(12 \alpha  \psi  \omega -27 \sqrt{2} r^{10}-15 \sqrt{2} r^9-20 r^6 \psi  \omega -6 r^5 \psi  \omega +15 \sqrt{2} \alpha  r^4\right) \right]  \right.\\
&  \left. \times  \frac{1}{r^2 \psi  \omega  \left(a^3+r^3\right)^3} \right\}\mathrm{d}r  = r_{+}^4 \left(2 a^3-r_{+}^3\right) \left[\frac{2 r_{+}^4 \omega  \left(\frac{\alpha_{r_{+}}^2 a_0}{r_{+}^4}\right){}^{3/2}}{a_0}  \right. \\
& \left. -  \sqrt{2} a_0 \left(-4 a^6 r_{+}^3 (8 r_{+}+3)+a^3 \left(\alpha_{r_{+}} -16 r_{+}^7-15 r_{+}^6+8 \alpha_{r_{+}}  r_{+}\right) \right.\right. \\
& \left.\left. +r_{+}^3 \left(\alpha_{r_{+}} -2 r_{+}^7-3 r_{+}^6+2 \alpha_{r_{+}}r_{+}\right)\right)\right] \\
& \times \frac{1}{2 \omega  \left(a^3+r_{+}^3\right)^2 \left(2 a^3 (4 r_{+}+3) r_{+}^3-\alpha_{r_{+}} +2 r_{+}^7+3 r_{+}^6-2 \alpha_{r_{+}} r_{+}\right) \sqrt{\frac{a_0 \left(a^3 (4 r_{+}+1) r_{+}^2-\alpha_{r_{+}} +r_{+}^6+r_{+}^5\right)}{r_{+}^4}}},
\end{split}
\fe
where,
\ie
\psi = \sqrt{\frac{a^3 (4 r+1)}{r^2}-\frac{\alpha }{r^4}+r^2+r},
\nonumber
\fe
\ie
\omega = \sqrt{4 a^3 r^3+r^6},
\nonumber
\fe
and
\ie
\alpha_{r_{+}} = \alpha(r_{+}).
\fe
After these considerations, all the \textit{Hawking} temperatures are agreement to each other, namely,
\ie
\overset{\nsim}{T}_{f}=T = \Tilde{T} = \Upsilon(r_{+},a) \Tilde{T}_{f}.
\fe
Therefore we can properly write the entropy $S$ as
\ie
S = \int \frac{\Upsilon(r_{+},a)}{\overset{\nsim}{T}_{f}} \mathrm{d} M_{-} = \pi r_{+}^{2} = \frac{A}{4}.
\fe

Another thermodynamic quantity worthy to be explored is the heat capacity. In this sense, we write
\ie
\begin{split}
C_{V} &= T \frac{\partial S}{\partial T} = T \frac{\partial S/\partial M}{\partial T/\partial M}\\
& = \frac{2 \pi M \left(\sqrt{a_0 M} \sqrt{4 a^3 r_{+}^3+r_{+}^6}+3 a_0 a^3 r_{+}\right)}{2 \sqrt{a_0 M} \sqrt{4 a^3 r_{+}^3+r_{+}^6}+3 a_0 a^3 r_{+}}\frac{\partial (r^{2}_{+})}{\partial M}.
\end{split}
\fe

\begin{figure}
    \centering
    \includegraphics[scale=0.437]{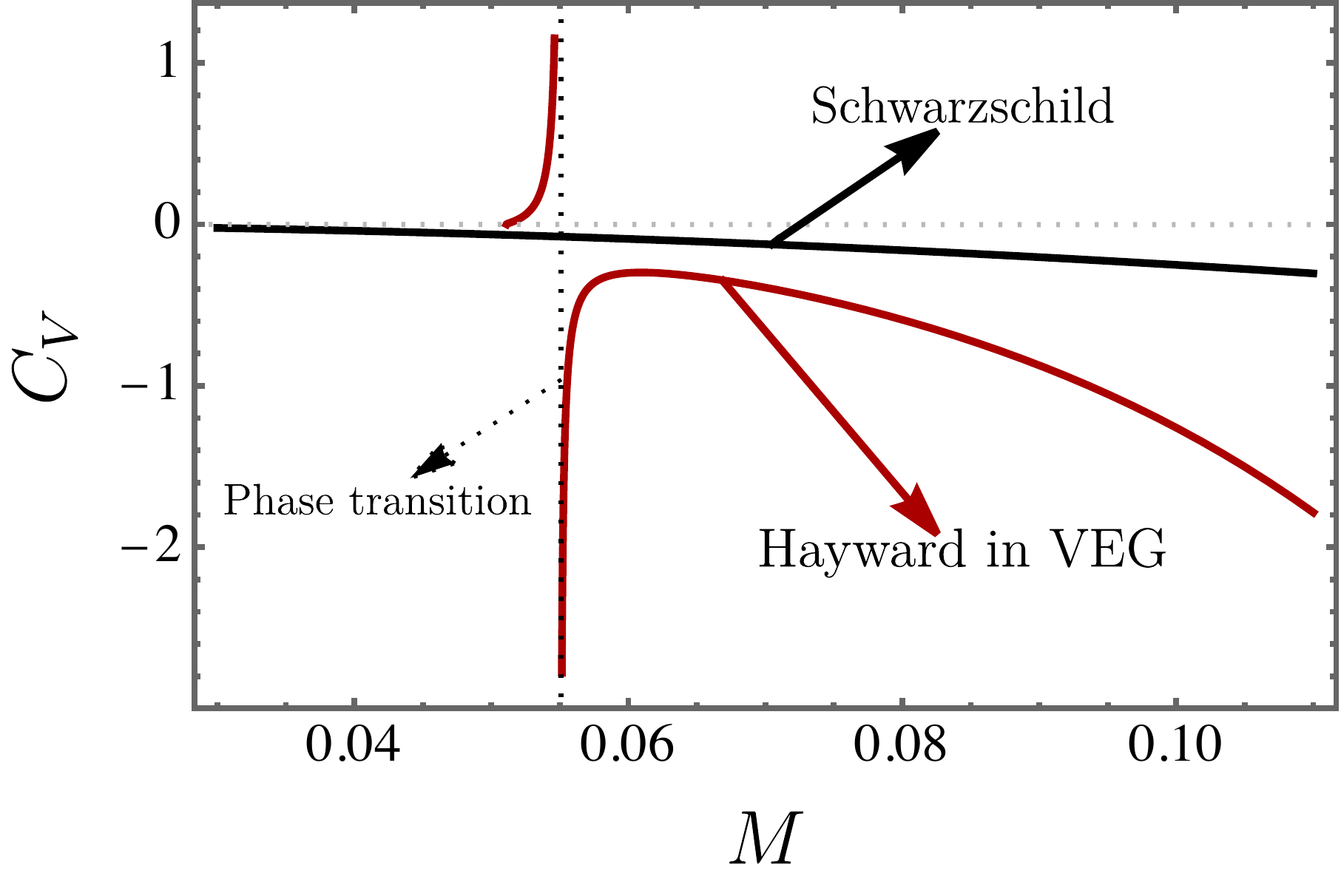}
    \includegraphics[scale=0.417]{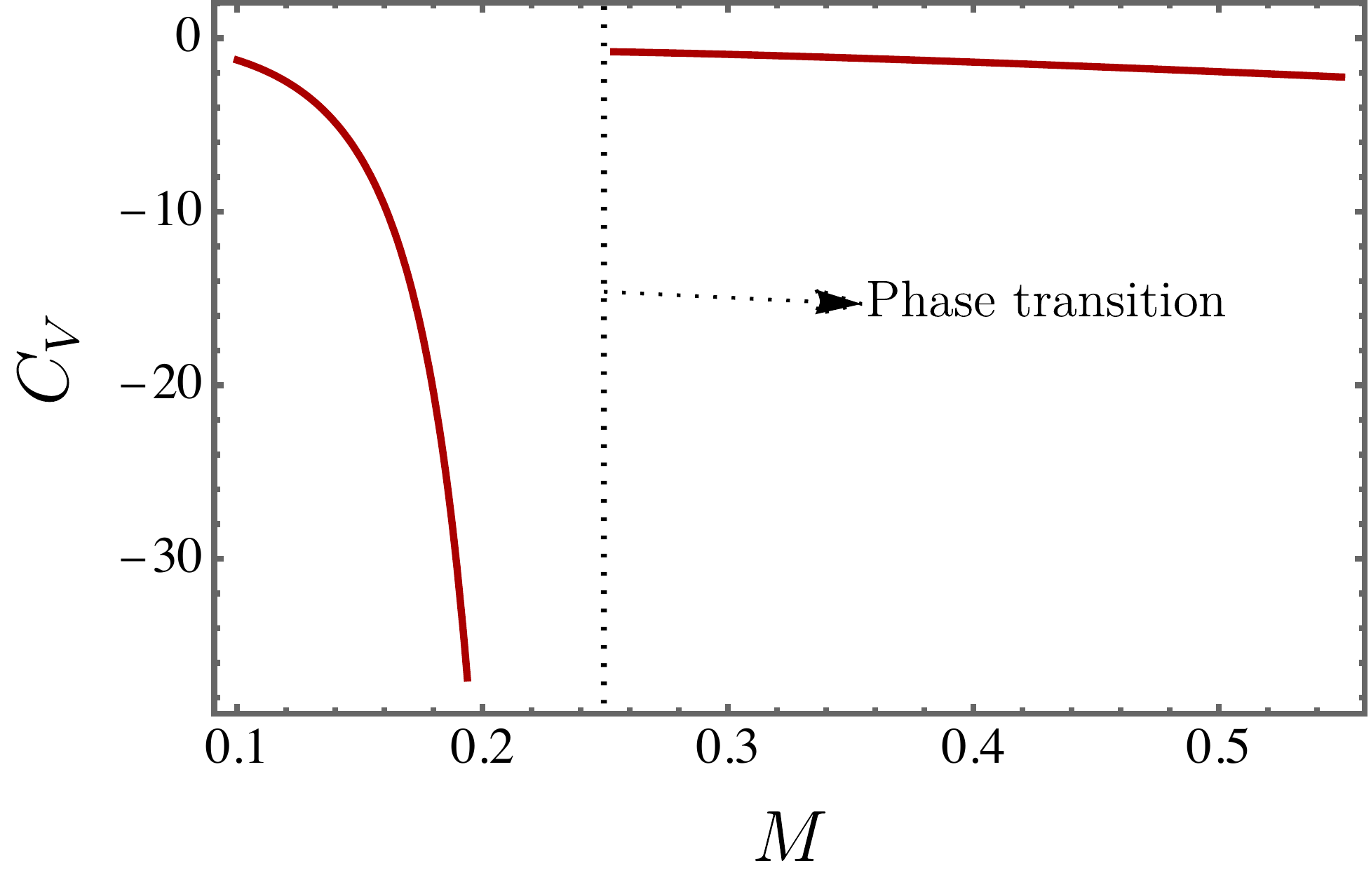}
    \caption{The heat capacity in comparison with the Schwarzschild case (on the left hand). A phase transition ($M \approx 0.548$) and a stable configuration ($M \approx 0.508-0.547$) are present. Also, it is highlighted another phase transition, i.e., $M \approx 0.249$.}
    \label{heatcapacity}
\end{figure}
In Fig. \ref{heatcapacity}, we exhibit the heat capacity in comparison with the Schwarzschild case (on the left side). In contrast, to the Hayward--like solution, it is also shown a discontinuity, being in agreement with the analysis made via the \textit{Hawking} temperature. In general, $g_{\mu\nu}$ has instabilities analogously to the Schwarzschild black hole. Nevertheless, there is a restrict range of mass which stability gives rise to instead, i.e., $M \approx 0.0508-0.0547$. In addition, as it is seen (on the right side) of the figure,  there exists also a phase transition in $M \approx 0.249$. This remarks are only possible due to existence of exotic matter. Furthermore, it is worth highlighting that the thermodynamic aspects have been extensively studied in various contexts, including cosmological scenarios \cite{araujo2021bouncing,campos2022quasinormal,maluf2018thermodynamics,araujo2022thermal,araujo2023thermodynamics,aa2,anacleto2021noncommutative,anacleto2021noncommutative1,aa4,aa6,aa7,aa10,aa13,aa14,aa15,filho2022thermodynamics} and other related areas \cite{aa11,anacleto2018lorentz,aa12,aguirre2021lorentz,aa1,qq}.

\begin{table}[t]
	\centering
		\caption{\label{thermo}Comparison of the thermodynamic properties between the Schwarzschild and Hayward--like black holes.}
	\setlength{\arrayrulewidth}{0.3mm}
	\setlength{\tabcolsep}{30pt}
	\renewcommand{\arraystretch}{1}
	\begin{tabular}{c c c}
		\hline \hline
		~ & Schwarzschild & Hayward in VEG  \\ \hline
	    \,\,\,$T$ & $1/8\pi M$ & $\frac{r_{+} \left(r_{+}^3-2 a^3\right) \left(M \sqrt{4 a^3 r_{+}^3+r_{+}^6}+3 a^3 r_{+} \sqrt{a_0 M}\right)}{\pi  \left(a^3+r_{+}^3\right)^2 \sqrt{4 a^3 r_{+}^3+r_{+}^6}}$ \\
		\,\,$A$ & $16 \pi M^{2} $ &$4\pi r_{+}^{2}$  \\ \
		$S$ &  $4 \pi M^{2}$ & $\pi r_{+}^{2}$ \\ 
		\,\,\,\,\,$C_{V}$ &  $- 8\pi {M^2}$ & $ \frac{2 \pi M \left(\sqrt{a_0 M} \sqrt{4 a^3 r_{+}^3+r_{+}^6}+3 a_0 a^3 r_{+}\right)}{2 \sqrt{a_0 M} \sqrt{4 a^3 r_{+}^3+r_{+}^6}+3 a_0 a^3 r_{+}}\frac{\partial (r^{2}_{+})}{\partial M}$ \\ \hline\hline
	\end{tabular}
\end{table}

\section{Geodesics}

\begin{figure}
    \centering
    \includegraphics[scale=0.33]{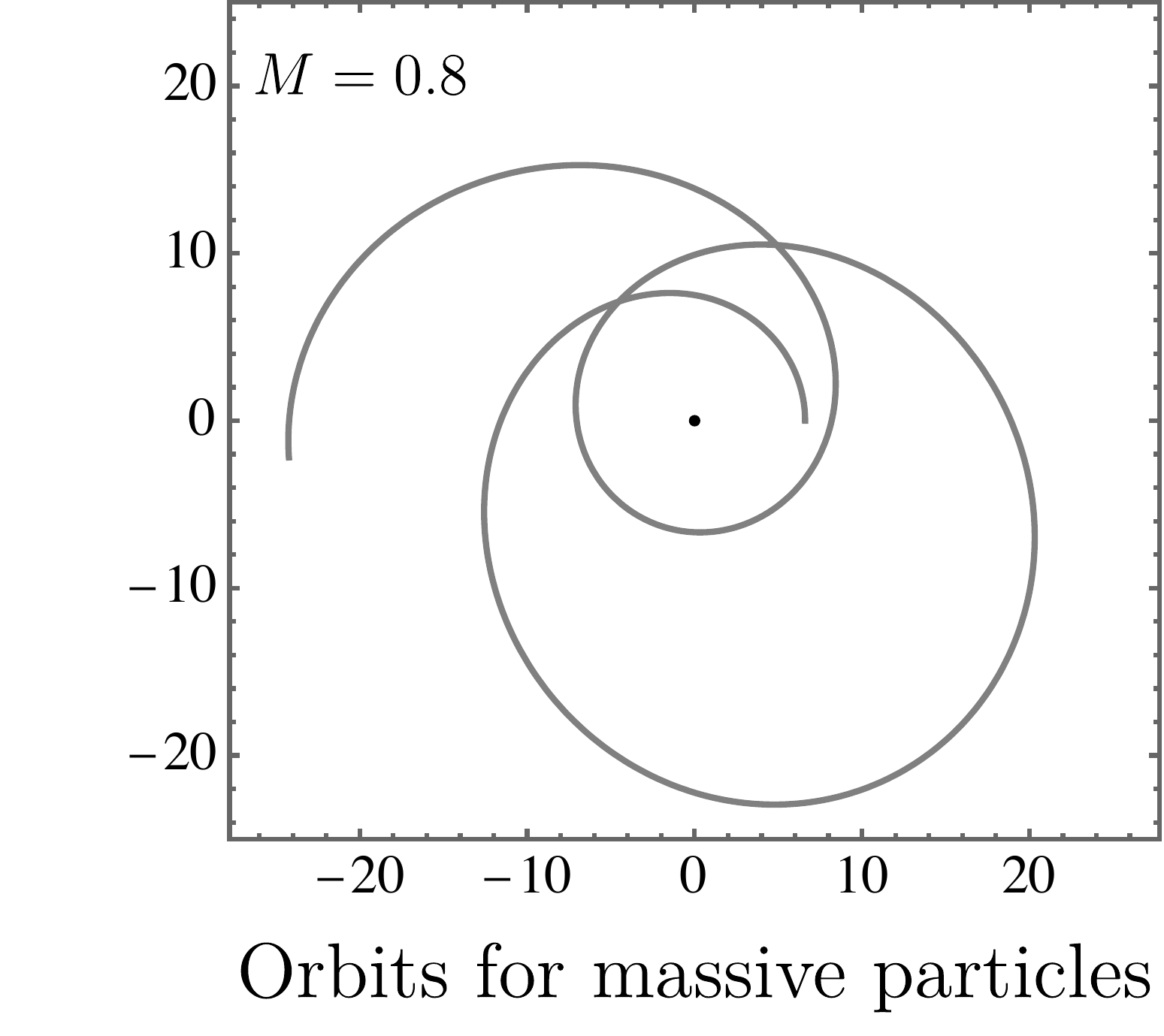}
    \includegraphics[scale=0.33]{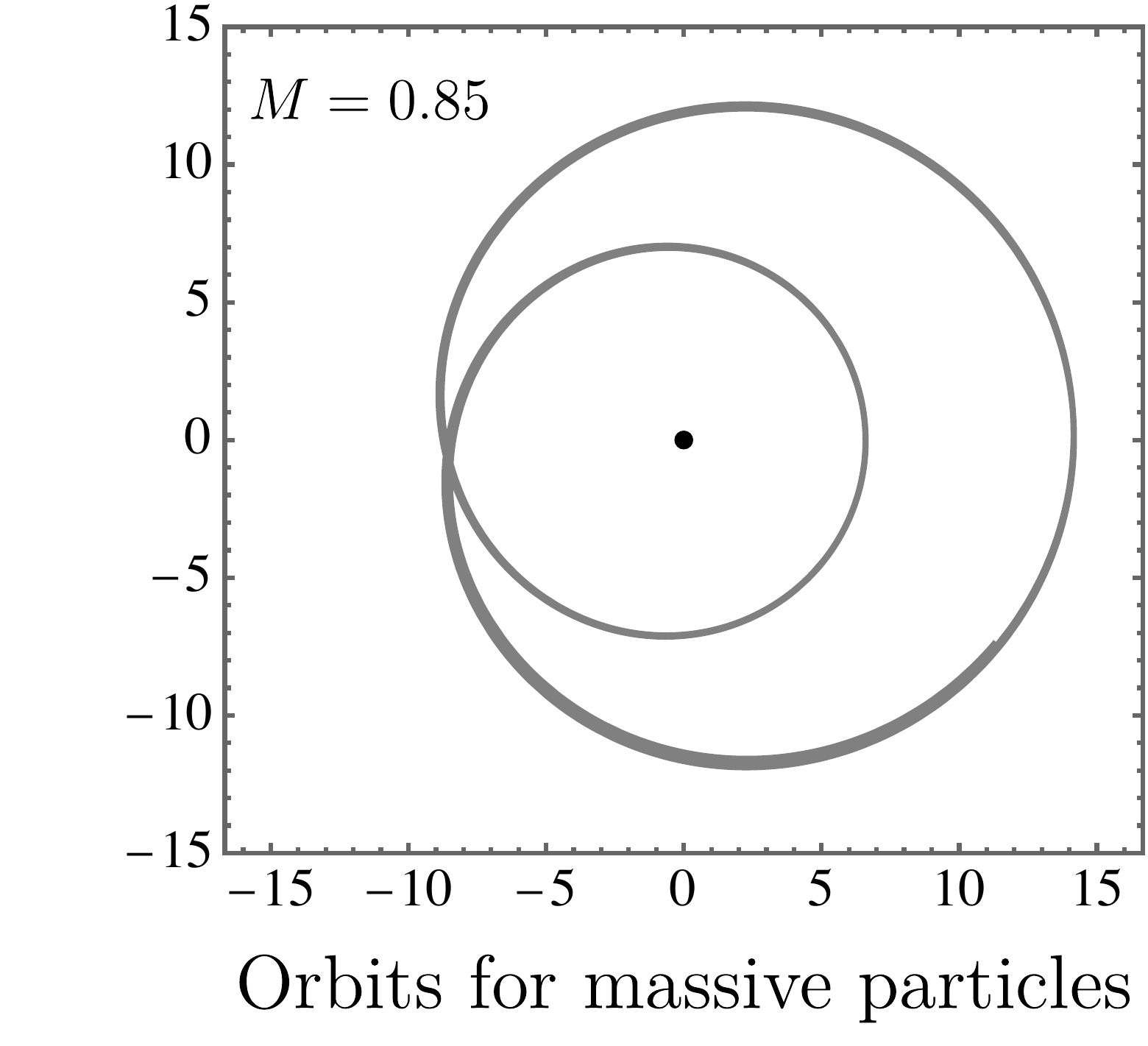}
    \includegraphics[scale=0.33]{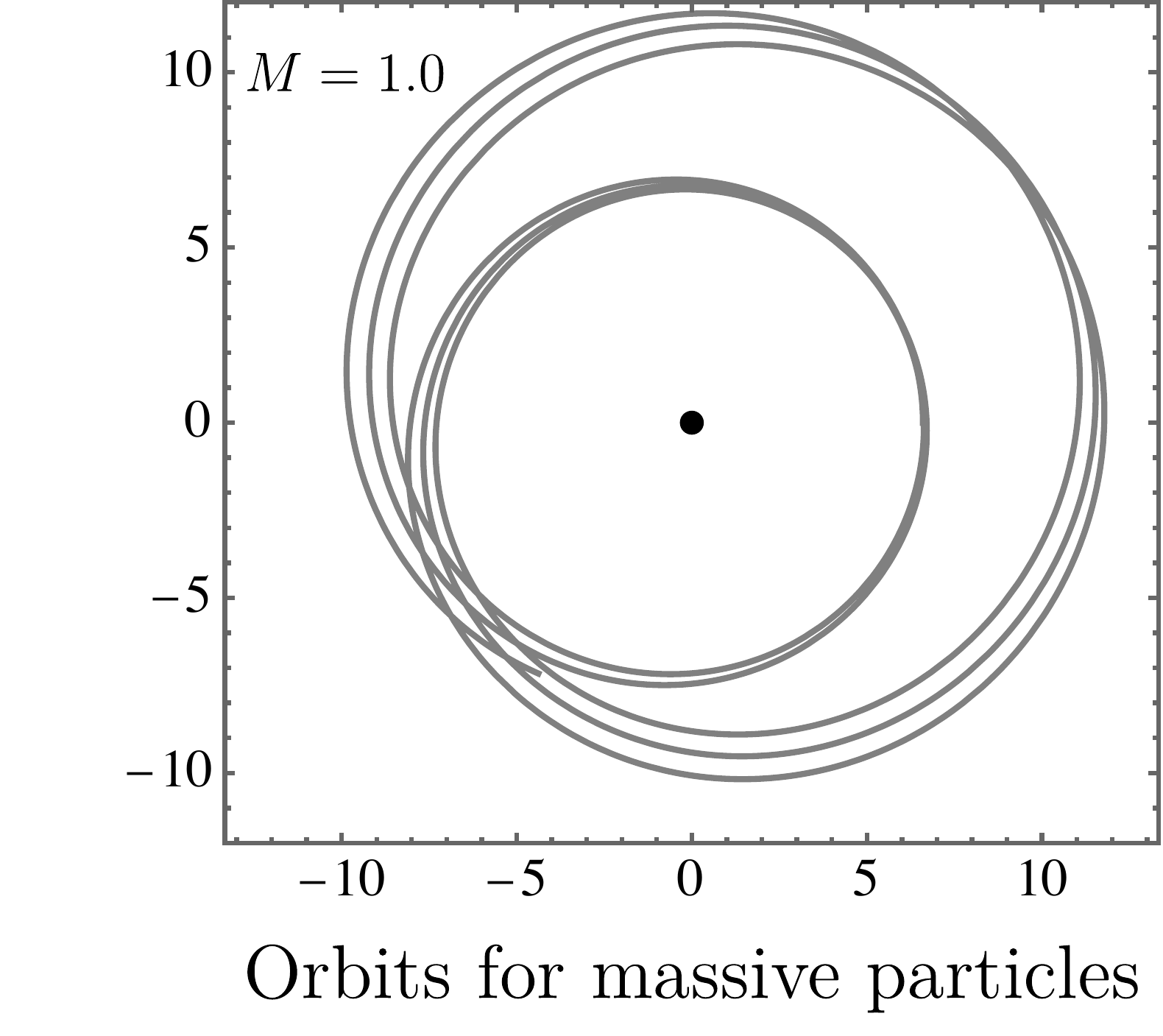}
    \caption{The trajectories of massive particles are shown for three distinct values of $M$. The black dot represents the event horizon $r_{+}$.}
    \label{deflectionoofmassiveparticles}
\end{figure}

\begin{figure}
    \centering
    \includegraphics[scale=0.4]{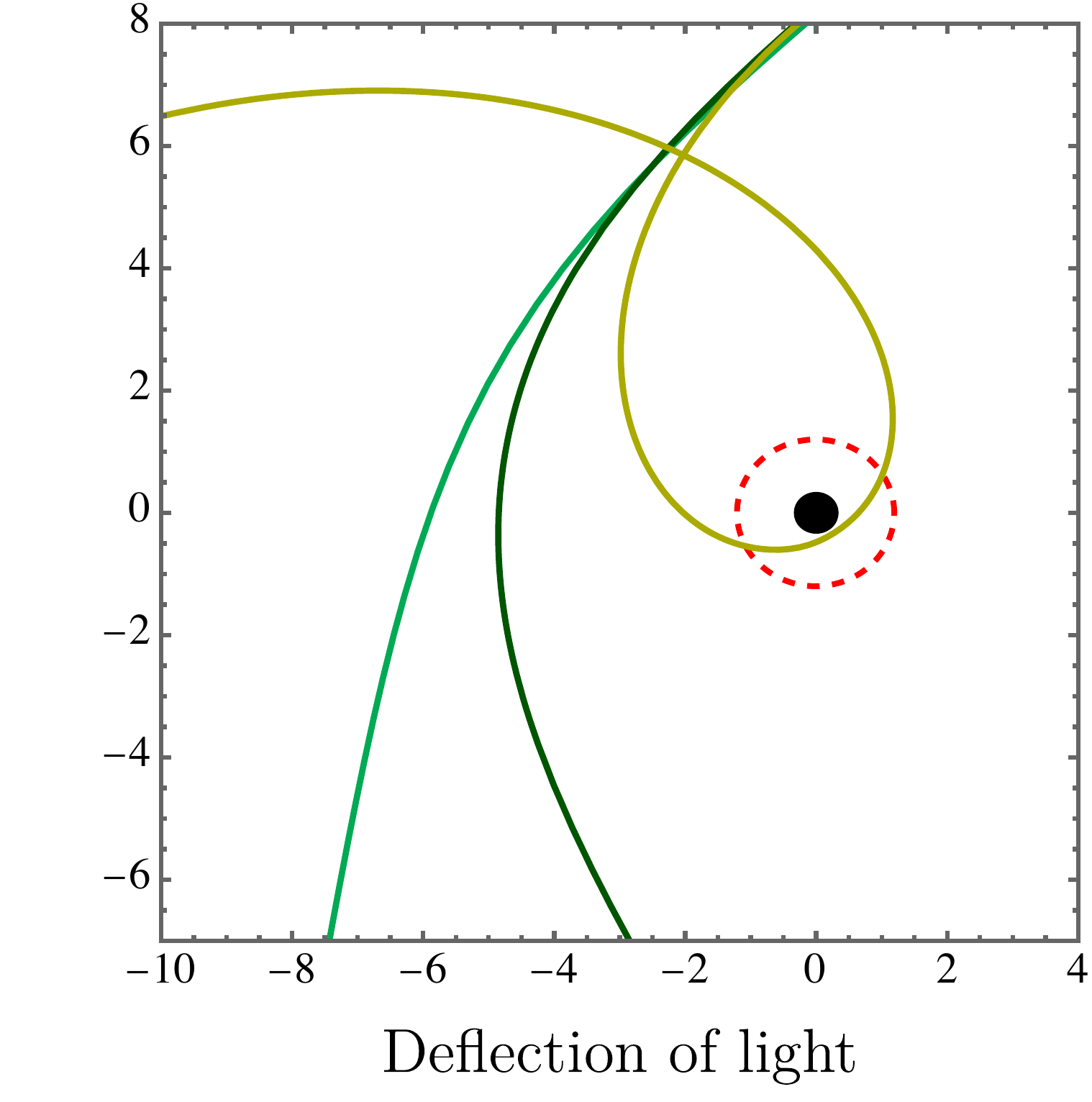}
    \caption{The trajectories for the light are shown. Red dashed lines represent the outer photon sphere (to $M=1$). The black dot represents the event horizon $r_{+}$. }
    \label{lightpath}
\end{figure}

The motion of particles in the context of emergent gravity has attracted considerable attention due to its potential implications for this theoretical framework \cite{lim2018field,verlinde2011origin}. Understanding the geodesic structure of Hayward--like black holes is particularly important for addressing astrophysical theoretical knowledge involving these objects, such as the properties of accretion disks and shadows. In other words, we shall focus on the complete behavior derived from the geodesic equation. To do so, we write
\ie
\frac{\mathrm{d}^{2}x^{\mu}}{\mathrm{d}s^{2}} + \Gamma\indices{^\mu_\alpha_\beta}\frac{\mathrm{d}x^{\alpha}}{\mathrm{d}s}\frac{\mathrm{d}x^{\beta}}{\mathrm{d}s} = 0, \label{geodesicfull}
\fe
where it gives rise to four coupled partial differential equations as follows:
\ie
\begin{split}
&\frac{\mathrm{d}}{\mathrm{d}s}t'(s) = -\frac{2 \sqrt{a_{0}M} r(s) \left(r(s)^3-2 a^3\right) r(s)' t(s)' \left(\sqrt{a_{0}M} \sqrt{4 a^3 r(s)^3+r(s)^6}+3 a^3 r(s)\right)}{\left(a^3+r(s)^3\right) \sqrt{4 a^3 r(s)^3+r(s)^6} \left(-2 \sqrt{a_{0}M} \sqrt{4 a^3 r(s)^3+r(s)^6}+a^3-2 M r(s)^2+r(s)^3\right)},
\end{split}
\fe
\ie
\begin{split}
\frac{\mathrm{d}}{\mathrm{d}s}r(s)' &= \frac{r(s) \left(\theta(s)'\right)^2 \left(-2 \sqrt{a_{0}M} \sqrt{4 a^3 r(s)^3+r(s)^6}+a^3-2 M r(s)^2+r(s)^3\right)}{a^3+r(s)^3}\\
& + \frac{r(s) \sin^2(\theta(s) ) \left(\varphi(s)'\right)^2 \left(-2 \sqrt{a_{0}M} \sqrt{4 a^3 r(s)^3+r(s)^6}+a^3-2 M r(s)^2+r(s)^3\right)}{a^3+r(s)^3}\\
& - \frac{1}{\left(a^3+r(s)^3\right)^3 \sqrt{4 a^3 r(s)^3+r(s)^6}} \left[  \left(-2 \sqrt{a_{0}M} \sqrt{4 a^3 r(s)^3+r(s)^6}+a^3-2 M r(s)^2+r(s)^3\right) \right. \\
& \left. \times \left(\sqrt{a_{0}M} \sqrt{4 a^3 r(s)^3+r(s)^6}+3 a^3 r(s)\right)\sqrt{a_{0}M} r(s) \left(r(s)^3-2 a^3\right) t(s)'^2 \right]  \\
& + \frac{\sqrt{a_{0}M} r(s) \left(r(s)^3-2 a^3\right) \left(r(s)'\right)^2 \left(\sqrt{a_{0}M} \sqrt{4 a^3 r(s)^3+r(s)^6}+3 a^3 r(s)\right)}{\left(a^3+r(s)^3\right) \sqrt{4 a^3 r(s)^3+r(s)^6} \left(-2 \sqrt{a_{0}M} \sqrt{4 a^3 r(s)^3+r(s)^6}+a^3-2 M r(s)^2+r(s)^3\right)},
\end{split}
\fe
\ie
\frac{\mathrm{d}}{\mathrm{d}s}\theta'(s) = \sin [\theta(s)]\cos [\theta(s) ] \varphi '(s)^2-\frac{2 \theta(s)' r(s)'}{r(s)},
\fe
\ie
\frac{\mathrm{d}}{\mathrm{d}s}\varphi'(s) = -\frac{2 \varphi(s)' \left(r(s)'+r(s) \theta(s)' \cot[ \theta(s)]\right)}{r(s)}.
\fe
As we can notice, the geodesic equation results four lengthy coupled partial differential equations. Based on these ones, the behavior of massive particles is displayed in Fig. \ref{deflectionoofmassiveparticles}. Here, we observe that as the mass increases, there is a greater ``compactification" of the massive trajectories. In other words, this phenomenon signifies that as the mass of an object grows, its trajectory becomes more compressed or concentrated, reflecting the intensified gravitational effects exerted by massive bodies. On the other hand, in Fig. \ref{lightpath}, we exhibit some different light paths.

\subsection{Critical orbits}

Understanding the dynamics of particles and the behavior of light rays near black hole structures requires a thorough grasp of the critical orbits. In our case, these orbits play a crucial role in unraveling the properties of spacetime influenced by dark matter effects.

In order to gain a deeper understanding of the impact on the photon sphere (also known as the critical orbit) of our black hole, we shall employ the Lagrangian method to calculate null geodesics. This method offers a more accessible approach for readers to comprehend the calculations, as compared to using the geodesic equation, as presented previously. By analyzing the influence of the black hole's mass, on the photon sphere, we can glean additional information regarding the gravitational effects of the Hayward--like solution in the Verlinde's emergent gravity approach. Such insights might hold potential implications for future observational astronomy.

Now, let us express the Lagrangian method as follows:
\begin{equation}
\mathcal{L} = \frac{1}{2} g_{\mu\nu}\Dot{x}^{\mu}\Dot{x}^{\nu},
\end{equation}
and when considering a fixed angle of $\theta=\pi/2$, the aforementioned expression simplifies to:
\ie
g_{00}^{-1} E^{2} + g_{11}^{-1} \Dot{r}^{2} + g_{33}L^{2} = 0,
\fe
with $E$ being the energy and $L$ the angular momentum. Above expression may be rewritten as
\ie
\Dot{r}^{2} = E^{2} - \left(  1 -\frac{2 M r^2}{a^3+r^3} - \frac{2 \sqrt{ M} \sqrt{r^3 \left(4 a^3+r^3\right)}}{a^3+r^3} \right)\left(  \frac{L^{2}}{r^{2}} \right),
\label{phtoton}
\fe
with $\overset{\nsim}{V} \equiv \left(  1 -\frac{2 M r^2}{a^3+r^3} - \frac{2 \sqrt{ M} \sqrt{r^3 \left(4 a^3+r^3\right)}}{a^3+r^3} \right)\left(  \frac{L^{2}}{r^{2}} \right)$ is the effective potential. To determine the critical radius, we need to solve the equation $\partial \overset{\nsim}{V}/\partial r = 0$. In order to obtain analytical solutions, we consider the regime that $a$ is small. Interestingly, Eq. (\ref{phtoton}) can yield until three different physical solutions under this configuration. This means that within the context of Hayward--like solution in Verlinde's emergent gravity, there may exist three critical orbits (or three photon spheres). Multiple photon spheres has also been recently reported in the literature \cite{guerrero2022multiring}. We will denote $r_{c_{-}}$, $r_{c_{\asymp}}$, and $r_{c_{+}}$ as as the inner, the middle and the outer photon spheres, respectively.

In Table \ref{criticalradius}, we present a summary of how variations in the parameters $M$ and $a$ influence critical orbits. When we increase the value of $M$ while keeping $a$ constant (specifically, $a=0.1$), both $r_{c_{-}}$ and $r_{c_{+}}$ exhibit an increase in their sizes. On the other hand, the behavior of $r_{\asymp}$ deviates from this trend; as we vary the mass values, its magnitude tends to decrease. In addition, a remarkable feature is worthy noting: for a large value of mass, i.e., $M=10^{9}$, $r_{c_{-}}$ and $r_{\asymp}$ will coincide. After this value, no matter how huge is the mass values, such a superposition is maintained.
Nonetheless, with an increase in the parameter $a$ (while maintaining $M$ at a constant value of $1$), we observe that both $r_{c_{-}}$ and $r_{c_{\asymp}}$ tend to exhibit higher values, in contrast to the behavior of $r_{c_{+}}$, which decreases.

\begin{table}[!h]
\begin{center}
\begin{tabular}{c c c c c} 
 \hline\hline
 \!$M$ \,\, $a$ & $r_{c_{-}}$ & $r_{c_{\asymp}}$ & $r_{c_{+}}$ \\ [0.2ex] 
 \hline 
 1.0 \, 0.1 & 0.120061 & 0.137208  & 0.99629 \\ 

 2.0 \,  0.1 & 0.121433 & 0.133082 & 1.56565 \\
 
 3.0 \, 0.1 & 0.122128 & 0.13152 & 2.01511  \\
 
 4.0 \, 0.1 & 0.122570 & 0.130653 & 2.39931 \\
 
 5.0 \, 0.1 & 0.122884 & 0.130087 & 2.74062 \\
 
 6.0 \, 0.1 & 0.123123 & 0.129682 & 3.05094 \\
 
 7.0 \, 0.1 & 0.123312 & 0.129374 & 3.33747 \\
 
 8.0 \, 0.1 & 0.123467 & 0.129131 & 3.60499 \\
 
 9.0 \, 0.1 & 0.123598 & 0.128931 & 3.85686 \\
 
 10.0  0.1 & 0.123709 & 0.128765 & 4.09556 \\
 [0.2ex] 
 \hline \hline
  \hline
  \!$M$ \,\, $a$ & $r_{c_{-}}$ & $r_{c_{\asymp}}$ & $r_{c_{+}}$ \\ [0.2ex] 
 \hline 
 1.0 \, 0.10 & 0.120061 & 0.137208  & 0.996299 \\ 

 1.0 \, 0.11 & 0.131442 & 0.152398 & 0.995058 \\
 
 1.0 \, 0.12 & 0.142714 & 0.167917 & 0.993559  \\
 
 1.0 \, 0.13 & 0.153879 & 0.183787 & 0.991773 \\
 
 1.0 \, 0.14 & 0.164937 & 0.200032 & 0.989669 \\
 
 1.0 \, 0.15 & 0.175890 & 0.216680 & 0.987213 \\
 
 1.0 \, 0.16 & 0.186739 & 0.233762 & 0.984366 \\
 
 1.0 \, 0.17 & 0.197485 & 0.251316 & 0.981088 \\
 
 1.0 \, 0.18 & 0.208128 & 0.269383 & 0.977329 \\
 
 1.0 \, 0.19 & 0.218670 & 0.288015 & 0.973036 \\

 1.0 \, 0.20 & 0.229112 & 0.307272 & 0.968143 \\
 [0.2ex] \hline \hline \hline
\end{tabular}
\caption{\label{criticalradius} The values of the critical orbits $r_{c_{-}}$, $r_{c_{\asymp}}$, and $r_{c_{+}}$ are shown for different values of mass $M$ and parameter $a$.}
\end{center}
\end{table}


\section{Shadows}

To investigate the effects of in Hayward--like solution in Verlinde's emergent gravity in the shadows, we regard
\begin{equation}\label{action1}
\frac{{\partial \mathcal{S}}}{{\partial \tau }} =  - \frac{1}{2}{g^{\mu \nu }}\frac{{\partial S}}{{\partial {\tau ^\mu }}}\frac{{\partial S}}{{\partial {\tau ^\nu }}}
\end{equation}
where the Jacobi action is denoted by $\mathcal{S}$, and an arbitrary affine parameter by $\tau$. Additionally, we can decompose such an expression as:
\begin{equation}\label{action2}
\mathcal{S} = \frac{1}{2}{m^2}\tau  - Et + L\phi  + {S_r}(r) + {S_\theta }(\theta ).
\end{equation}
We can express $S_r(r)$ and $S_{\theta}(\theta)$ as mathematical functions that are dependent on the variables $r$ and $\theta$, respectively. Considering our focus on the behavior of photons, it follows that the energy $E$ and angular momentum $L$ become conserved quantities. By employing Eqs. (\ref{action1}) and (\ref{action2}), we can derive the equations that govern the trajectory of the photon, which are commonly referred to as the null geodesic equations
\begin{equation}\label{time}
\frac{{\mathrm{d}t}}{{\mathrm{d}\tau }} = \frac{E}{f(r)},
\end{equation}
\begin{equation}\label{position}
\frac{{\mathrm{d}r}}{{\mathrm{d}\tau }} = \frac{{\sqrt {\mathcal{R}(r)} }}{{{r^2}}},
\end{equation}
\begin{equation}\label{thetadot}
\frac{{\mathrm{d}\theta }}{{\mathrm{d}\tau }} =\pm \frac{{\sqrt {\mathcal{Q} (r)} }}{{{r^2}}},
\end{equation}
\begin{equation}\label{phidot}
\frac{{\mathrm{d}\varphi }}{{\mathrm{d}\tau }} = \frac{{L\,{{\csc }^2}\theta }}{{{r^2}}},
\end{equation}
where $\mathcal{R}(r)$ and $\mathcal{Q}(\theta)$ are 
\begin{equation}
\mathcal{R}(r) = {E^2}{r^4} - (\mathcal{K} + {L^2}){r^2}f(r)
\end{equation}
\begin{equation}
\mathcal{Q} (\theta ) = \mathcal{K} - {L^2}{\cot ^2}\theta.
\end{equation}
Here, the symbol $\mathcal{K}$ refers to the Carter constant.
Eq. (\ref{thetadot}) contains plus and minus signs, which represent the photon's motion in the outgoing and ingoing radial directions, respectively. To simplify the analysis without loss of generality, we set the angle $\theta$ to $\pi/2$ and focus on the equatorial plane. By adopting this simplification, our focus shifts to the radial equation, in which we introduce the notion of an effective potential $\mathcal{V}_{eff}(r)$
\begin{equation}\label{veff2}
{\left(\frac{{\mathrm{d}r}}{{\mathrm{d}\tau }}\right)^2} + {\mathcal{V}_{eff}}(r) = 0,
\end{equation}
where, it reads
\begin{equation}\label{veffr}
\mathcal{V}_{eff}(r) = (L^2+\mathcal{K})\frac{f(r)}{{{r^2}}} - {E^2},
\end{equation}
and we conveniently introduce two new parameters to accomplish our analysis, namely,
\begin{equation}\label{par}
\xi  = \frac{L}{E}
\text{ and }
\eta  = \frac{\mathcal{K}}{{E^2}}.
\end{equation}
In addition, by using Eq. (\ref{veffr}) and Eq. (\ref{par}), we have 
\begin{equation}
{\xi ^2} + \eta  =  \frac{\tilde{r}^{2}_{c}}{{f(\tilde{r}_{c})}},
\end{equation}
where $\tilde{r}_{c}$ represents our critical orbits displayed in the previous section.

To determine the radius of the shadow, we will make use of the celestial coordinates $\alpha$ and $\beta$ \cite{singh2018shadow}, which are related to the constants of motion as follows: $\alpha=-\xi$ and $\beta=\pm\sqrt{\eta}$. By employing these coordinates, we can express the shadow radius as:

\begin{equation}
\mathcal{R}_{\text{Shadow}} = \frac{\tilde{r}_{c}}{{\sqrt{|f(\tilde{r}_{c})|}}}.
\end{equation}

\begin{figure}
    \centering
    \includegraphics[scale=0.345]{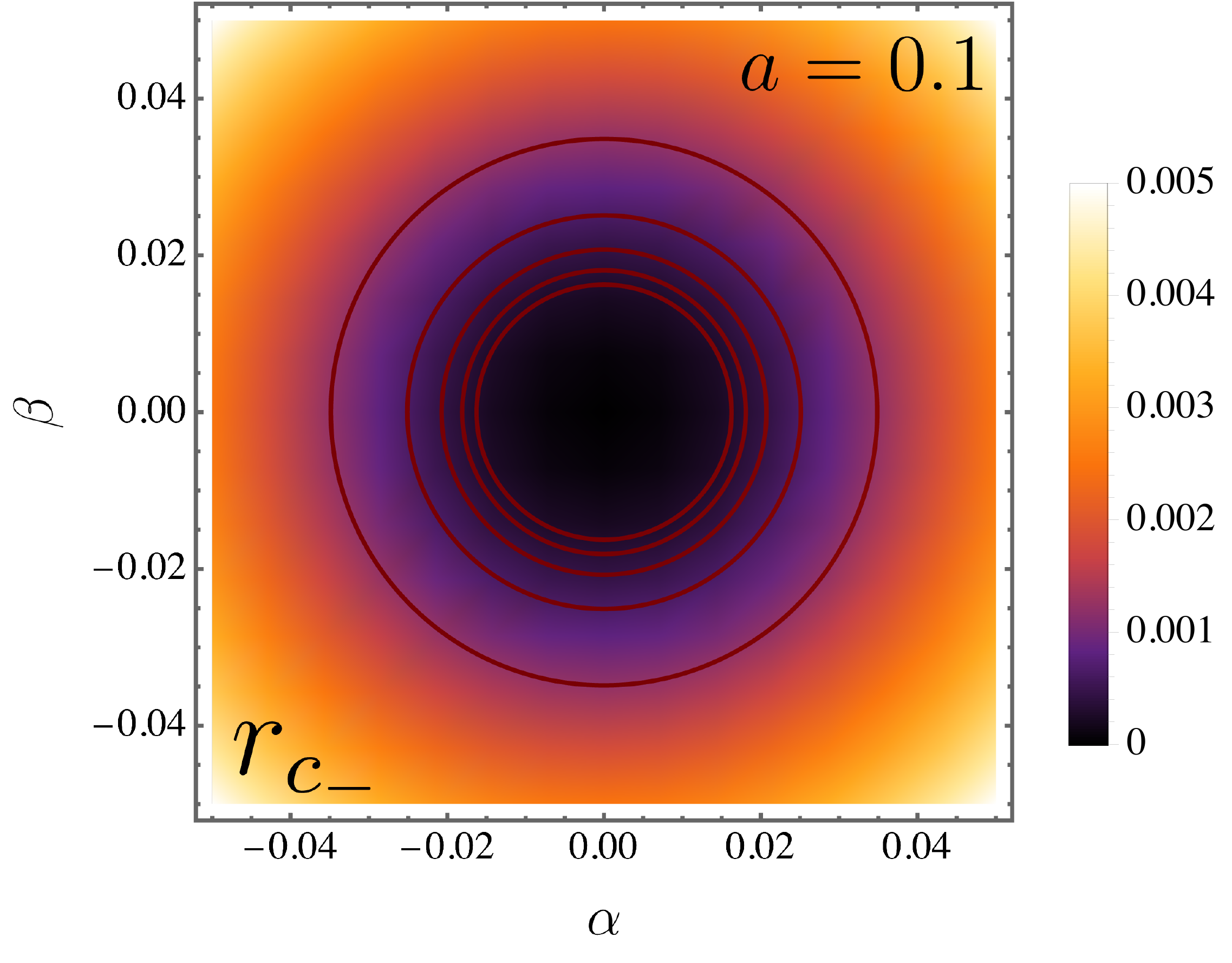}
    \includegraphics[scale=0.355]{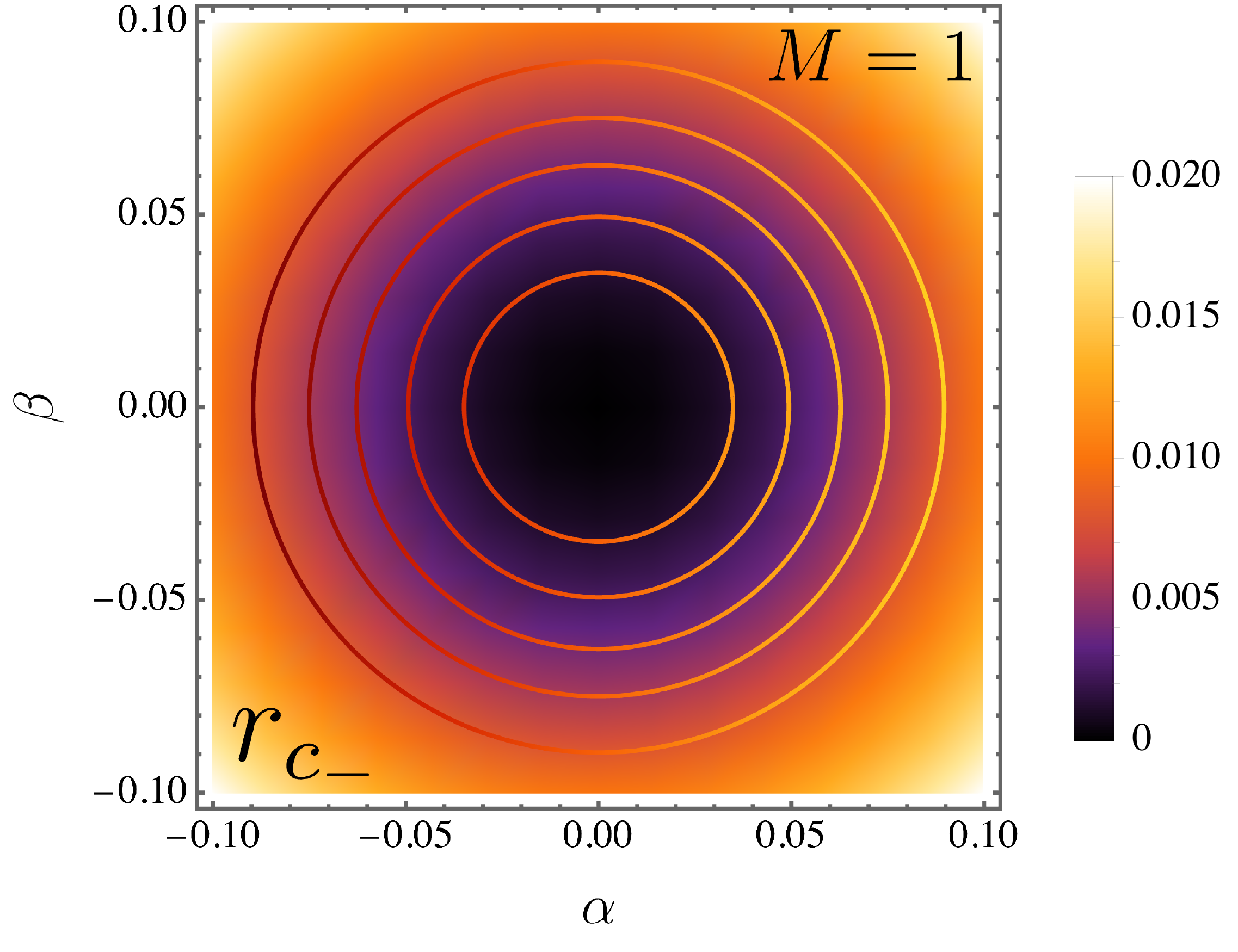}
    \caption{On the left--hand side, we explore the impact of varying the parameter $M$ with specific values, namely, $M=1, 2, 3, 4, 5$. Notably, the outermost radius corresponds to $M=1$. It is evident that an increase in mass directly leads to a reduction in the shadow radius. On the right--hand side, we examine the shadow radius for different values of the parameter $a$ while maintaining a constant mass of $M=1$. In this context, as the parameter $a$ increases, represented by values such as $a=0.100, 0.125, 0.150, 0.175, 0.200$, the size of the shadows proportionally expands. }
    \label{shadows}
\end{figure}

\begin{figure}
    \centering
    \includegraphics[scale=0.35]{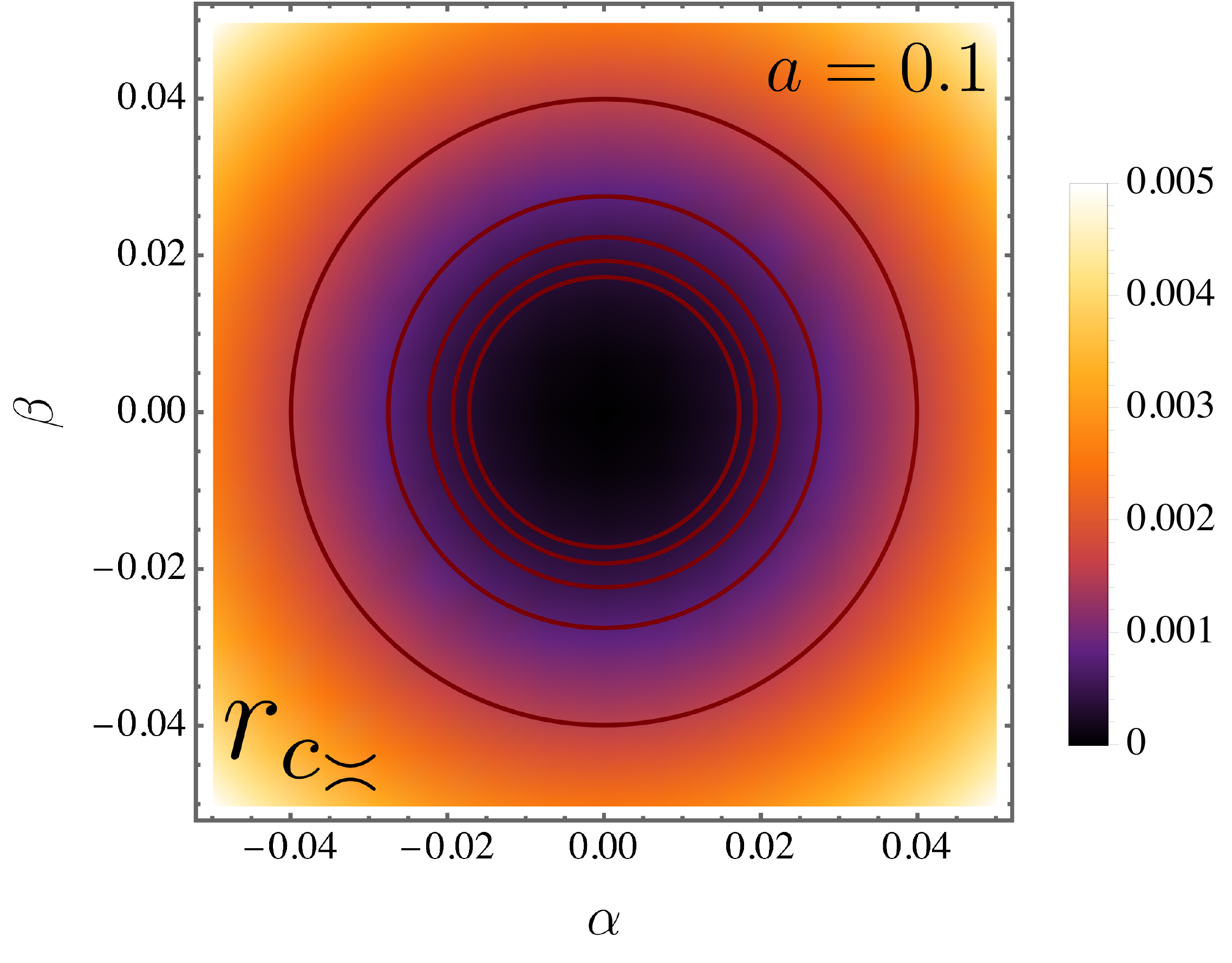}
    \includegraphics[scale=0.35]{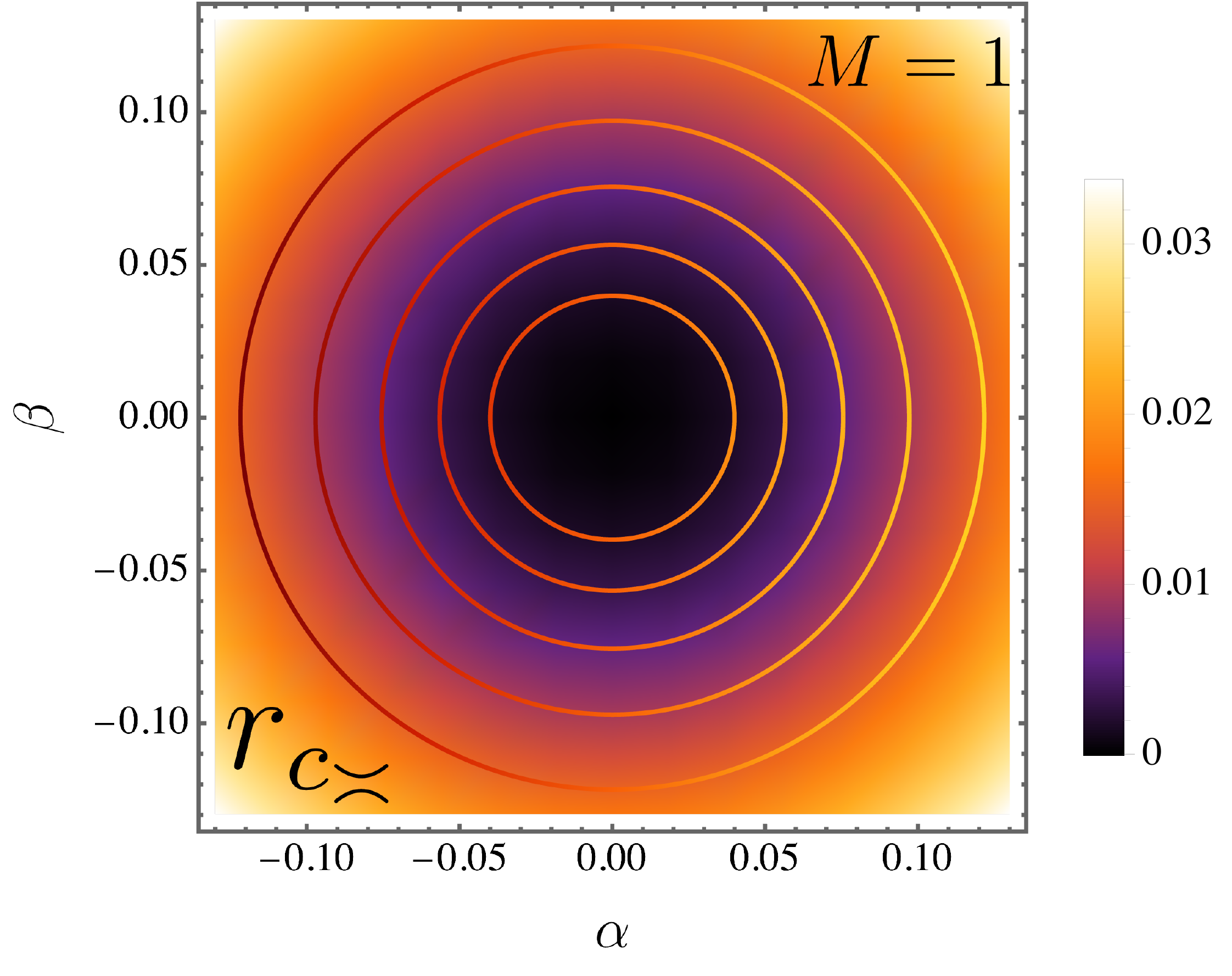}
    \caption{On the left side of our analysis, we investigate the effects of varying the parameter $M$ using specific values: $M=1, 2, 3, 4, 5$. It is worth noting that the largest radius corresponds to $M=1$. Clearly, an increase in mass directly results in a reduction of the shadow radius. On the right side of our examination, we assess the shadow radius under different values of the parameter $a$, while keeping the mass constant at $M=1$. In this context, as the parameter $a$ increases, represented by values such as $a=0.100, 0.125, 0.150, 0.175, 0.200$, the size of the shadows correspondingly expands. }
    \label{shadows2}
\end{figure}

Figures \ref{shadows} and \ref{shadows2} presents an analysis of the shadows produced by our black hole corresponding to the photon sphere $r_{c_{-}}$ and $r_{c_{\asymp}}$, respectively, with variations in the parameters $M$ and $a$. On the left side, we illustrate the effect of varying $M$, specifically with values of $M=1, 2, 3, 4, 5$. Notably, the outermost radius corresponds to $M=1$. As evident, an increase in the mass directly contributes to the reduction in the shadow radius. On the right side, we showcase the shadow radius under different values of the parameter $a$, while keeping the mass constant at $M=1$. In this context, as the value of parameter $a$ rises, i.e., $a=0.100, 0.125, 0.150, 0.175, 0.200$, so does the size of the shadows.

\begin{figure}
    \centering
    \includegraphics[scale=0.35]{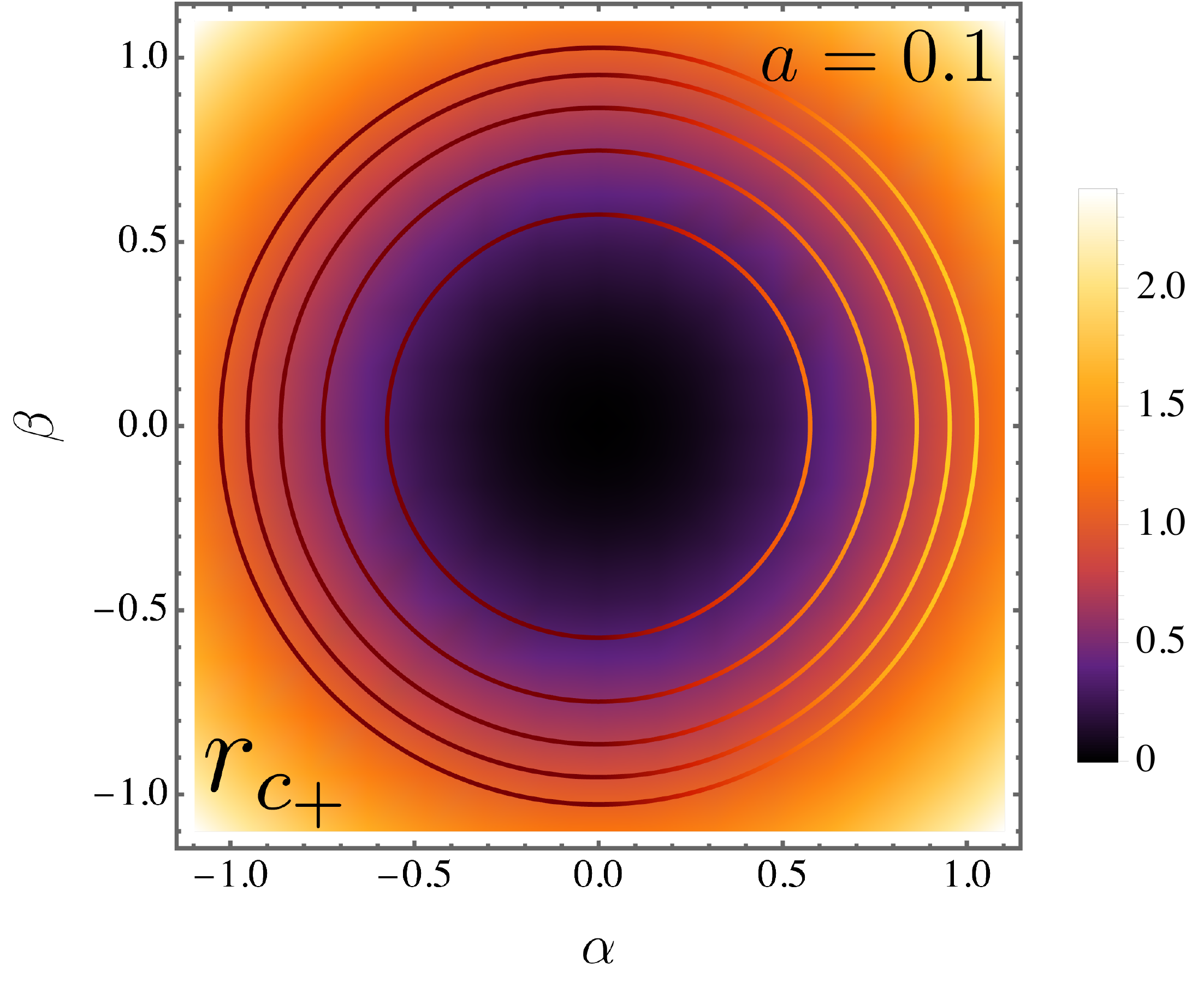}
    \includegraphics[scale=0.35]{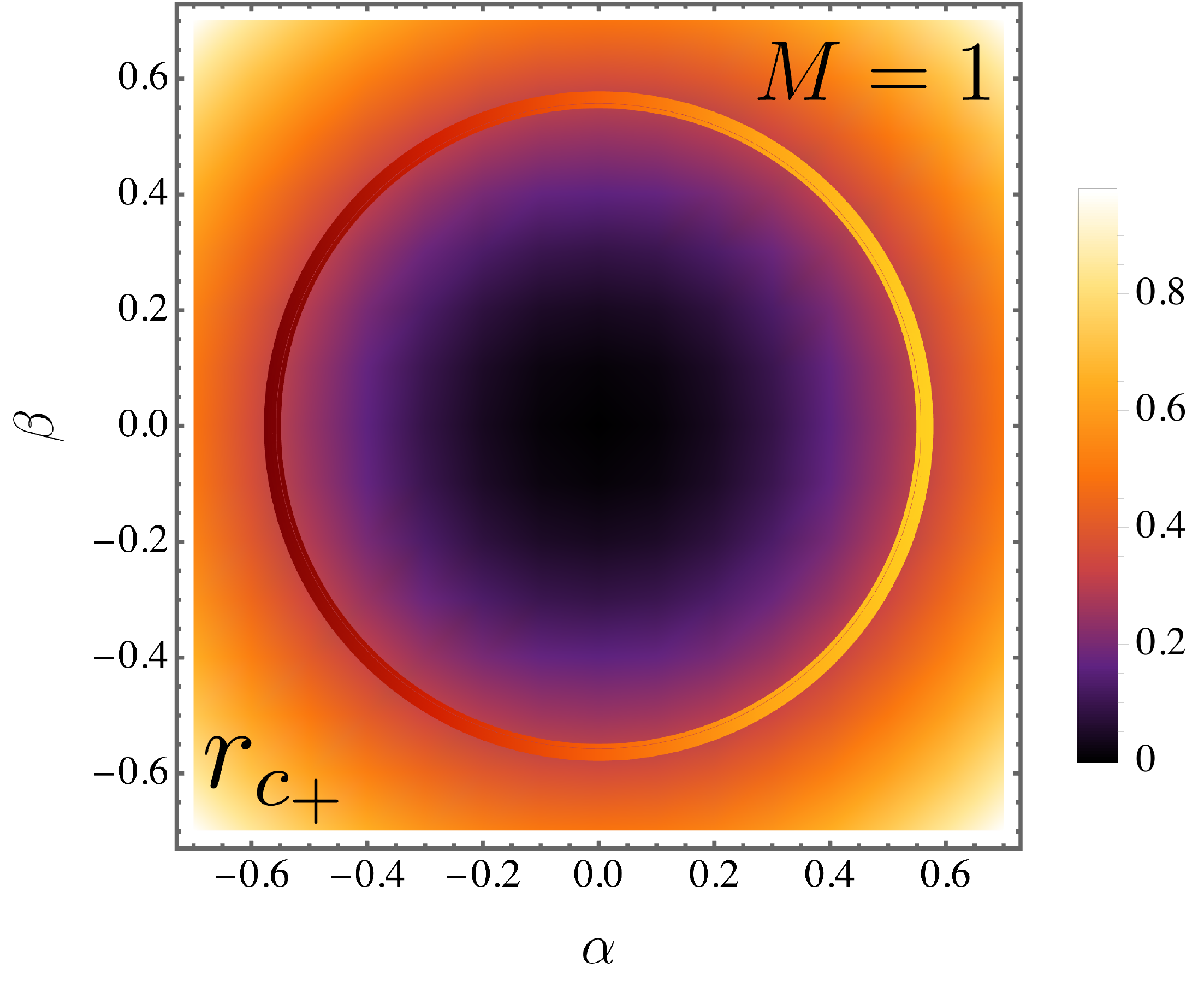}
    \caption{On the left side of our analysis, we undertake an examination of the influence of varying the parameter $M$ using specific values: $M=1, 2, 3, 4, 5$. It is noteworthy that the largest radius corresponds to $M=1$. Evidently, an increase in mass directly leads to an expansion of the shadow radius. On the right side of our investigation, we scrutinize the shadow radius for varying values of the parameter $a$, all while maintaining a constant mass of $M=1$. In this context, as the parameter $a$ increases, as denoted by values such as $a=0.100, 0.125, 0.150, 0.175, 0.200$, the dimensions of the shadows correspondingly contract.}
    \label{shadows3}
\end{figure}

Furthermore, when considering the shadows associated with $r_{c_{+}}$, we maintain the same parameters as in the previous analyses conducted for $r_{c_{-}}$ and $r_{c_{\asymp}}$. However, what we observe is precisely the opposite behavior: as the mass increases, the shadows also increase, whereas when the parameter $a$ is augmented, the shadows decrease. These dynamics are prominently illustrated in Figure \ref{shadows3}. Additionally, a subtle discrepancy is evident in the shadow radii to the latter case. This variation is anticipated since the shadows of a black hole are inherently linked to the associated critical orbits, aligning with the findings detailed in Table \ref{criticalradius} for $r_{c_{+}}$.

It is worth mentioning that the reduction of the photon sphere, as a function of parameters which govern the black holes, was also recently reported in the literature in the context of a black hole coupled with nonlinear electrodynamics, considering different values of the magnetic charge and the deviation constant  \cite{singh2022thermodynamic}. To our case, our parameter $a$ seems to ``mimic" the effects brought about by these ones. Additionally, a similar behavior is seen for the shadows as well \cite{singh2022quasinormal}.


\section{The quasinormal modes}

During the ringdown phase, a captivating phenomenon referred to as \textit{quasinormal} modes becomes apparent, showcasing distinct oscillation patterns that remain unaffected by the initial perturbations. These modes reflect the intrinsic properties of the system and are the result of the natural oscillations of spacetime itself, independent of specific initial conditions.

In contrast to \textit{normal} modes, which are associated with closed systems, \textit{quasinormal} modes are linked to open systems. As a consequence, these modes gradually dissipate energy through the emission of gravitational waves. Mathematically, they can be described as poles of the complex Green function.

To determine their frequencies, we need to find solutions to the wave equation within a system described by a background metric $g_{\mu\nu}$. However, obtaining analytical solutions for such modes is generally a formidable task.

\begin{figure}
    \centering
    \includegraphics[scale=0.25]{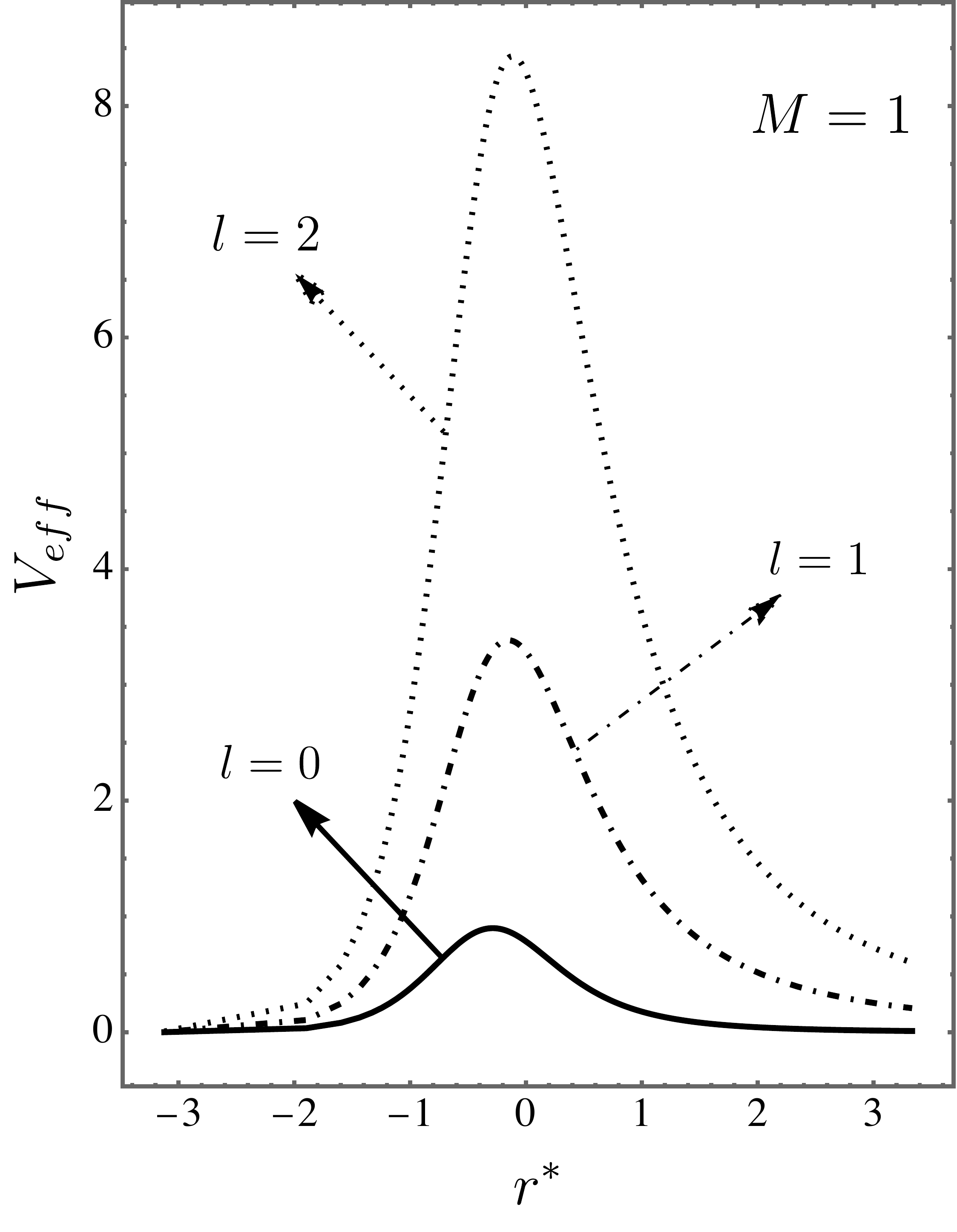}
    \caption{The effective potential $V_{eff}$ as a function of the tortoise coordinate $r^{*}$ for different values of $l$.}
    \label{effectivepotential}
\end{figure}

Several techniques have been proposed in the scientific literature to obtain solutions for them. Among these methods, the WKB (Wentzel--Kramers--Brillouin) approach stands out as one of the most used. Its development can be traced back to the groundbreaking work of Will and Iyer \cite{iyer1987black,iyer1987black1}, and subsequent advancements up to the sixth order were made by Konoplya \cite{konoplya2003quasinormal}. For our calculations, we specifically focus on analyzing perturbations using the scalar field. Consequently, we write the Klein--Gordon equation within the context of a curved spacetime
\ie
\frac{1}{\sqrt{-g}}\partial_{\mu}(g^{\mu\nu}\sqrt{-g}\partial_{\nu}\Phi) = 0.\label{KL}
\fe
Although the investigation of \textit{backreaction} effects in this specific scenario is captivating, this manuscript does not delve into this aspect and instead focuses on other aspects. Our primary focus is directed towards studying the scalar field as a small perturbation. Moreover, due to the presence of spherical symmetry, we can take advantage of this opportunity to decompose the scalar field in a specific manner, as detailed below:
\ie
\Phi(t,r,\theta,\varphi) = \sum^{\infty}_{l=0}\sum^{l}_{m=-l}r^{-1}\Psi_{lm}(t,r)Y_{lm}(\theta,\varphi),\label{decomposition}
\fe
where by expressing the spherical harmonics as $Y_{lm}(\theta,\varphi)$, we can substitute the decomposition of the scalar field, as indicated in Eq. (\ref{decomposition}), into Eq. (\ref{KL}). This substitution leads us to a Schrödinger--like equation. Thereby, this one exhibits wave--like properties, rendering it suitable for our analysis
\ie
-\frac{\partial^{2} \Psi}{\partial t^{2}}+\frac{\partial^{2} \Psi}{\partial r^{*2}} + V_{eff}(r^{*})\Psi = 0.\label{schordingereq}
\fe
The potential $V_{eff}$ is commonly known as the \textit{Regge--Wheeler} potential or the effective potential. It encapsulates valuable information about the geometry of the black hole. Furthermore, we introduce the tortoise coordinate $r^{*}$, which extends over the entire spacetime as $r^{*}\rightarrow \pm \infty$. After conducting several algebraic manipulations, we can explicitly express the effective potential as 
\ie
\begin{split}
V_{eff}(r) = f(r) \left[\frac{-\frac{4 M r}{a^3+r^3}+\frac{6 M r^4}{\left(a^3+r^3\right)^2}+\frac{6 r^2 \sqrt{a_0 M} \sqrt{r^3 \left(4 a^3+r^3\right)}}{\left(a^3+r^3\right)^2}-\frac{\sqrt{a_0 M} \left(3 r^2 \left(4 a^3+r^3\right)+3 r^5\right)}{\left(a^3+r^3\right) \sqrt{r^3 \left(4 a^3+r^3\right)}}}{r}+\frac{l (l+1)}{r^2}\right]
\end{split}.
\fe
In Fig. \ref{effectivepotential}, we display the effective potential $V_{eff}$ as a function of the tortoise coordinate $r^{*}$, considering different values of $l$.

\subsection{The WKB method}

In this section, our goal is to obtain stationary solutions, which are defined by assuming that $\Psi(t,r)$ can be expressed as $\Psi(t,r) = e^{-i\omega t} \psi(r)$, where $\omega$ represents the frequency. By making this assumption, we can conveniently separate the time--independent component of Eq. (\ref{schordingereq}) using the following approach:
\ie
\frac{\partial^{2} \psi}{\partial r^{*2}} - \left[  \omega^{2} - V_{eff}(r^{*})\right]\psi = 0.\label{timeindependent}
\fe
To solve Eq. (\ref{timeindependent}), it is essential to consider the relevant boundary conditions. In our specific case, the solutions that fulfill the required conditions are characterized by pure ingoing behavior near the horizon
\[
    \psi^{\text{in}}(r^{*}) \sim 
\begin{cases}
    C_{l}(\omega) e^{-i\omega r^{*}} & ( r^{*}\rightarrow - \infty)\\
    A^{(-)}_{l}(\omega) e^{-i\omega r^{*}} + A^{(+)}_{l}(\omega) e^{+i\omega r^{*}} & (r^{*}\rightarrow + \infty).\label{boundaryconditions11}
\end{cases}
\]
The complex constants $C_{l}(\omega)$, $A^{(-)}_{l}(\omega)$, and $A^{(+)}_{l}(\omega)$ hold significant importance in the subsequent discussion. In general, the frequencies $\omega_{nl}$ that satisfy the condition $A^{(-)}_{l}(\omega_{nl})=0$ are referred to as the \textit{quasinormal} modes of a black hole. This condition ensures that the modes correspond to a purely outgoing wave at spatial infinity and a purely ingoing wave at the event horizon. Here, the integers $n$ and $l$ represent the overtone and multipole numbers, respectively. The spectrum of \textit{quasinormal} modes is characterized by the eigenvalues of Eq. (\ref{timeindependent}). To analyze these modes, we utilize the WKB method, a semi--analytical approach that leverages the analogy with quantum mechanics.

The WKB approximation, initially introduced by Schutz and Will \cite{schutz1985black}, has been applied to calculate the \textit{quasinormal} modes in the context of particle scattering around black holes. Subsequent advancements in this technique were made by Konoplya \cite{konoplya2003quasinormal, konoplya2004quasinormal}. However, it is important to note that for this method to be valid, the potential must exhibit a barrier--like shape, approaching constant values as $r^{*} \rightarrow \pm \infty$. By fitting the power series of the solution near the turning points of the maximum potential, the \textit{quasinormal} modes can be obtained \cite{santos2016quasinormal}. The Konoplya formula for calculating these modes is expressed as follows:
\ie
\frac{i(\omega^{2}_{n}-V_{0})}{\sqrt{-2 V^{''}_{0}}} - \sum^{6}_{j=2} \Lambda_{j} = n + \frac{1}{2}.
\fe

In the previously mentioned expression, Konoplya's formula for the \textit{quasinormal} modes incorporates several essential elements. The term $V^{''}_{0}$ represents the second derivative of the potential evaluated at its maximum point $r_{0}$, while $\Lambda_{j}$ denotes constants that rely on the effective potential and its derivatives at the maximum. It is worth emphasizing that recent progress in the field has introduced a 13th--order WKB approximation, proposed by Matyjasek and Opala \cite{matyjasek2017quasinormal}, which further enhances the accuracy of calculating the \textit{quasinormal} modes.

The following tables, namely Tables \ref{table5}, and \ref{table6}, depict a presentation of the \textit{quasinormal} frequencies for the third--order WKB method. On the other hand, in Tables \ref{table11} and \ref{table22}, we display such frequencies, being calculated via sixth--order WKB method instead.
The tables are categorized based on the multipole number $l$ and mass $M$. In addition, we see to the third-- and sixth--order approximations (when $l=0$), $\omega_{2}$ turns out to be unstable. Similarly, to the latter one, (when $l=1$), $\omega_{1}$, and $\omega_{2}$ present instabilities. Such a behavior might be a direct consequence of the dark matter introduced at the initial conditions to obtain a Hayward--like black hole solution.

It is of significant importance to note that the \textit{quasinormal} modes associated with the scalar field exhibit a negative imaginary part. This characteristic indicates that these modes undergo exponential decay over time, representing the dissipation of energy through scalar waves. This observation is consistent with previous studies investigating scalar, electromagnetic, and gravitational perturbations in spherically symmetric geometries \cite{berti2009quasinormal, konoplya2011quasinormal, heidari2023gravitational, chen2023quasinormal}.

Notice that, for both order of approximations employed here, there exists an increase of the real part, while the imaginary part decreases as $M$ runs. This observation implies that the mass parameter plays a crucial role in this scenario as it governs the damping behavior of the scalar waves. Depending on the specific value of $M$, the damping can occur at either a faster or slower rate.

\begin{table}[!h]
\begin{center}
\begin{tabular}{c c c c} 
 \hline\hline
 $M$ & $\omega_{0}$ & $\omega_{1}$ & $\omega_{2}$  \\ [0.2ex] 
 \hline 
 1.0 & 2.09010 - 2.14868$i$ & 1.11193 - 6.36919$i$ & \text{Unstable}  \\ 
 
 1.1 & 2.27535 - 2.30724$i$ & 1.13400 - 6.88540$i$ & \text{Unstable} \\
 
 1.2 & 2.45484 - 2.46099$i$ & 1.14985 - 7.39063$i$  & \text{Unstable}\\
 
 1.3 & 2.62931 - 2.61066$i$ & 1.16079 - 7.88652$i$  & \text{Unstable}\\
 
 1.4 & 2.79935 - 2.75683$i$ & 1.16783 - 8.37433$i$ & \text{Unstable} \\
 
 1.5 & 2.96545 - 2.89995$i$ & 1.17177 - 8.85507$i$ & \text{Unstable} \\
 
 1.6 & 3.12803 - 3.04040$i$ & 1.17321 - 9.32953$i$ & \text{Unstable} \\
 
 1.7 & 3.28742 - 3.17847$i$ & 1.17264 - 9.79835$i$ & \text{Unstable} \\
 
 1.8 & 3.44394 - 3.31443$i$ & 1.17047 - 10.2621$i$ & \text{Unstable} \\
  
 1.9 & 3.59784 - 3.44848$i$ & 1.16701 - 10.7211$i$ & \text{Unstable} \\
   
 2.0 & 3.74932 - 3.58081$i$ & 1.16250 - 11.1759$i$ & \text{Unstable} \\ [0.2ex] 
 \hline \hline
\end{tabular}
\caption{\label{table5}The \textit{quasinormal} frequencies by using third--order WKB approximation for different values of mass $M$. In this case, the multipole number is $l=0$.}
\end{center}
\end{table}

\begin{table}[!h]
\begin{center}
\begin{tabular}{c c c c} 
 \hline\hline
 $M$ & $\omega_{0}$ & $\omega_{1}$ & $\omega_{2}$  \\ [0.2ex] 
 \hline 
 1.0 & 1.10190 - 3.44389$i$ & 3.79108 - 9.19885$i$ & 8.44732 - 16.1494$i$  \\ 
 
 1.1 & 1.20922 - 3.59082$i$ & 3.30326 - 9.41880$i$ & 6.96713 - 16.0962$i$ \\
 
 1.2 & 1.33885 - 3.74433$i$ & 3.00024 - 9.74870$i$  & 5.86976 - 16.3740$i$\\
 
 1.3 & 1.47862 - 3.89923$i$ & 2.80379 - 10.1316$i$  & 5.01190 - 16.8360$i$\\
 
 1.4 & 1.62310 - 4.05337$i$ & 2.67240 - 10.5413$i$ & 4.31268 - 17.4052$i$ \\
 
 1.5 & 1.76962 - 4.20580$i$ & 2.58248 - 10.9647$i$ & 3.72265 - 18.0391$i$ \\
 
 1.6 & 1.91679 - 4.35616$i$ & 2.51989 - 11.3947$i$ & 3.21001 - 18.7127$i$ \\
 
 1.7 & 2.06383 - 4.50431$i$ & 2.47581 - 11.8274$i$ & 2.75358 - 19.4112$i$ \\
 
 1.8 & 2.21029 - 4.65026$i$ & 2.44450 - 12.2607$i$ & 2.33890 - 20.1255$i$ \\
  
 1.9 & 2.35593 - 4.79409$i$ & 2.42214 - 12.6932$i$ & 1.95583 - 20.8497$i$ \\
   
 2.0 & 2.50059 - 4.93588$i$ & 2.40613 - 13.1242$i$ & 1.59716 - 21.5800$i$ \\ [0.2ex] 
 \hline \hline
\end{tabular}
\caption{\label{table6}The \textit{quasinormal} frequencies by using third--order WKB approximation for different values of mass $M$. In this case, the multipole number is $l=1$.}
\end{center}
\end{table}

\begin{table}[!h]
\begin{center}
\begin{tabular}{c c c c} 
 \hline\hline
 $M$ & $\omega_{0}$ & $\omega_{1}$ & $\omega_{2}$  \\ [0.2ex] 
 \hline 
 1.0 & 1.83131 - 2.33411$i$ & 0.787900 - 6.14765$i$ & \text{Unstable}  \\ 
 
 1.1 & 2.00285 - 2.49760$i$ & 0.745305 - 6.59571$i$ & \text{Unstable} \\
 
 1.2 & 2.16962 - 2.65514$i$ & 0.688251 - 7.04092$i$  & \text{Unstable}\\
 
 1.3 & 2.33211 - 2.80768$i$ & 0.619818 - 7.48648$i$  & \text{Unstable}\\
 
 1.4 & 2.49077 - 2.95596$i$ & 0.542691 - 7.93441$i$ & \text{Unstable} \\
 
 1.5 & 2.64595 - 3.10056$i$ & 0.459158 - 8.38592$i$ & \text{Unstable} \\
 
 1.6 & 2.79797 - 3.24198$i$ & 0.371129 - 8.84157$i$ & \text{Unstable} \\
 
 1.7 & 2.94710 - 3.38062$i$ & 0.280164 - 9.30153$i$ & \text{Unstable} \\
 
 1.8 & 3.09357 - 3.51680$i$ & 0.187517 - 9.76563$i$ & \text{Unstable} \\
  
 1.9 & 3.23759 - 3.65080$i$ & 0.094172 - 10.2336$i$ & \text{Unstable} \\
   
 2.0 & 3.37935 - 3.78285$i$ & 0.000895 - 10.7049$i$ & \text{Unstable} \\ [0.2ex] 
 \hline \hline
\end{tabular}
\caption{\label{table11}The \textit{quasinormal} frequencies by using sixth--order WKB approximation for different values of mass $M$. The multipole number considered here is $l=0$.}
\end{center}
\end{table}

\begin{table}[!h]
\begin{center}
\begin{tabular}{c c c c} 
 \hline\hline
 $M$ & $\omega_{0}$ & $\omega_{1}$ & $\omega_{2}$ \\ [0.2ex] 
 \hline 
 1.0 & 0.43028 - 4.32521$i$ & \text{Unstable} & \text{Unstable} \\ 
 
 1.1 & 0.98794 - 3.83805$i$ & \text{Unstable} & \text{Unstable} \\
 
 1.2 & 1.26813 - 3.91498$i$ & \text{Unstable} & \text{Unstable} \\
 
 1.3 & 1.44216 - 4.10463$i$ & \text{Unstable} & \text{Unstable} \\
 
 1.4 & 1.58174 - 4.30962$i$ & \text{Unstable} & \text{Unstable} \\
 
 1.5 & 1.71011 - 4.50924$i$ & \text{Unstable} & \text{Unstable} \\
 
 1.6 & 1.83489 - 4.69988$i$ & \text{Unstable} & \text{Unstable} \\

 1.7 & 1.95865 - 4.88190$i$ & \text{Unstable} & \text{Unstable} \\
 
 1.8 & 2.08219 - 5.05642$i$ & \text{Unstable} & \text{Unstable} \\
 
 1.9 & 2.20572 - 5.22457$i$ & \text{Unstable} & \text{Unstable} \\
 
 2.0 & 2.32922 - 5.38727$i$ & \text{Unstable} & \text{Unstable} \\ [0.2ex] 
 \hline \hline
\end{tabular}
\caption{\label{table22}The \textit{quasinormal} frequencies by using sixth--order WKB approximation for different values of $M$. The multipole number is $l=1$.}
\end{center}
\end{table}


\section{Time--Domain solution}

To comprehensively examine the influence of the \textit{quasinormal} spectrum on time--dependent scattering phenomena, a meticulous investigation of scalar perturbations in the time domain becomes essential. However, due to the intricate nature of our effective potential, a more precise methodology is necessary to gain deeper understanding. In this regard, we employ the characteristic integration method, originally developed by Gundlach and collaborators \cite{gundlach1994late}, as a powerful tool to effectively study the problem. By utilizing this approach, we can attain valuable insights into the role of \textit{quasinormal} modes in time--dependent scattering scenarios, thereby significantly contributing to the exploration of black holes and associated phenomena.

The approach described in Ref. \cite{gundlach1994late} centers around the application of light--cone coordinates, denoted as $u = t - x$ and $v = t + x$. By employing these coordinates, the wave equation can be reformulated in a more appropriate manner, facilitating a thorough analysis of the system
\ie
\left(  4 \frac{\partial^{2}}{\partial u \partial v} + V(u.v)\right) \Psi (u,v) = 0 \label{timedomain}.
\fe
In order to achieve efficient integration of the aforementioned expression, a discretization scheme can be utilized, leveraging a straightforward finite--difference method and numerical techniques. By adopting this approach, the equation can be numerically integrated effectively, yielding precise and efficient results
\ie
\Psi(N) = -\Psi(S) + \Psi(W) + \Psi(E) - \frac{h^{2}}{8}V(S)[\Psi(W) + \Psi(E)] + \mathcal{O}(h^{4}),
\fe
where $S=(u,v)$, $W=(u+h,v)$, $E=(u,v+h)$, and $N=(u+h,v+h)$, with $h$ representing the overall grid scale factor. The null surfaces $u=u_{0}$ and $v=v_{0}$ hold particular importance as they serve as the designated locations for specifying initial data. In our study, we have opted to utilize a Gaussian profile centered at $v=v_{c}$, with a width of $\sigma$, which is chosen on the null surface $u=u_{0}$
\ie
\Psi(u=u_{0},v) = A e^{-(v-v_{*})^{2}}/2\sigma^{2}, \,\,\,\,\,\, \Psi(u,v_{0}) = \Psi_{0}.
\fe
At $v=v_{0}$, a constant initial condition $\Psi(u,v_{0}) = \Psi_{0}$ was imposed, and for the sake of simplicity, we assume $\Psi_{0} = 0$. Once the null data is specified, the integration process proceeds along the lines of constant $u$, in the direction of increasing $v$. 

In this study, we present the outcomes of our investigation on the scalar test field. To ensure convenience, the null data was defined by a Gaussian profile centered at $u = 10$ on the $u=0$ surface. The profile had a width of $\sigma = 3$, and we assumed $\Psi_{0}=0$. A grid was set up to encompass the intervals $u \in [0,20]$ and $v \in [0,20]$, utilizing grid points sampled to achieve an overall grid spacing of $h = 0.1$.

\begin{figure}
    \centering
    \includegraphics[scale=0.425]{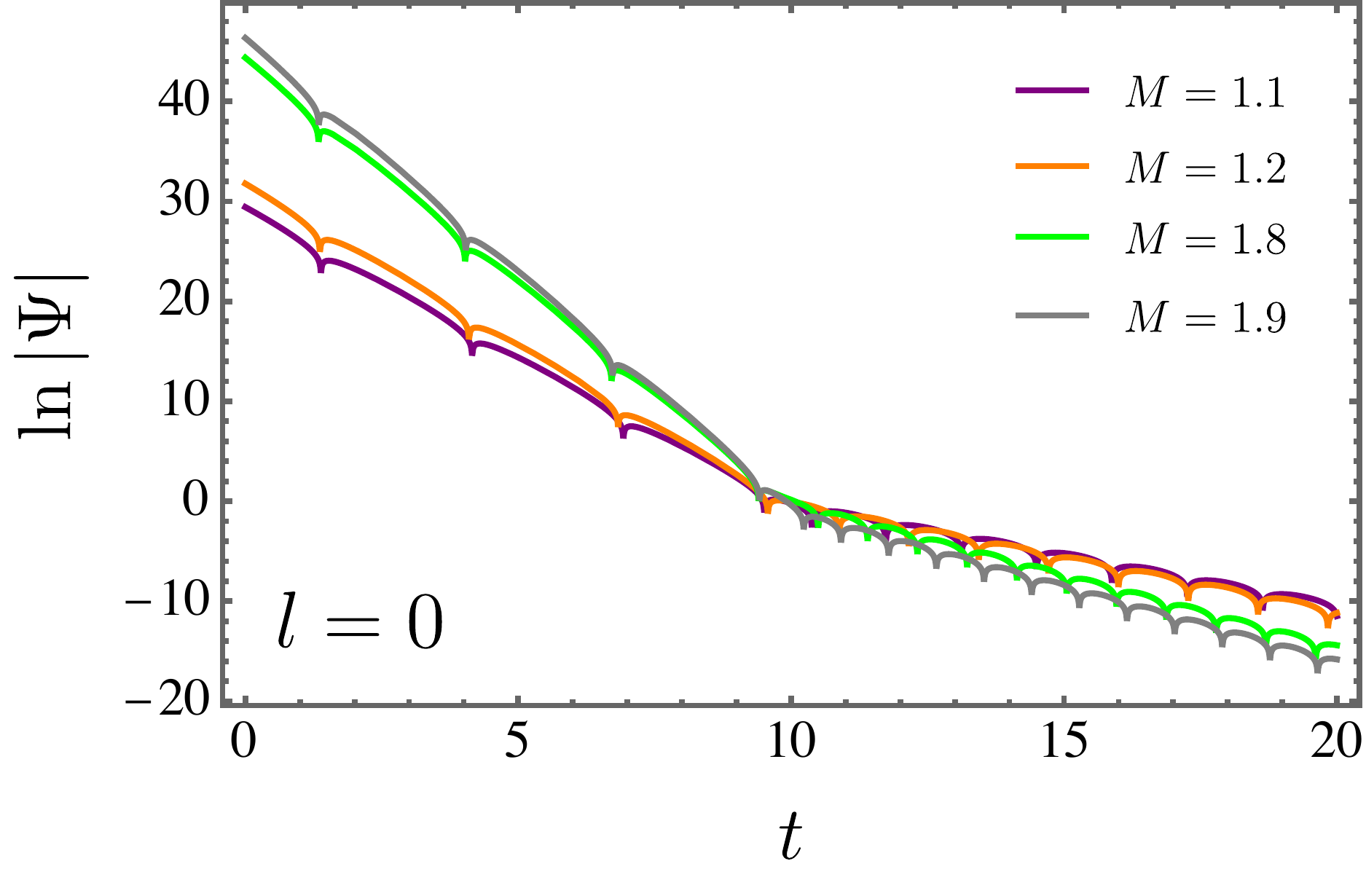}
    \includegraphics[scale=0.425]{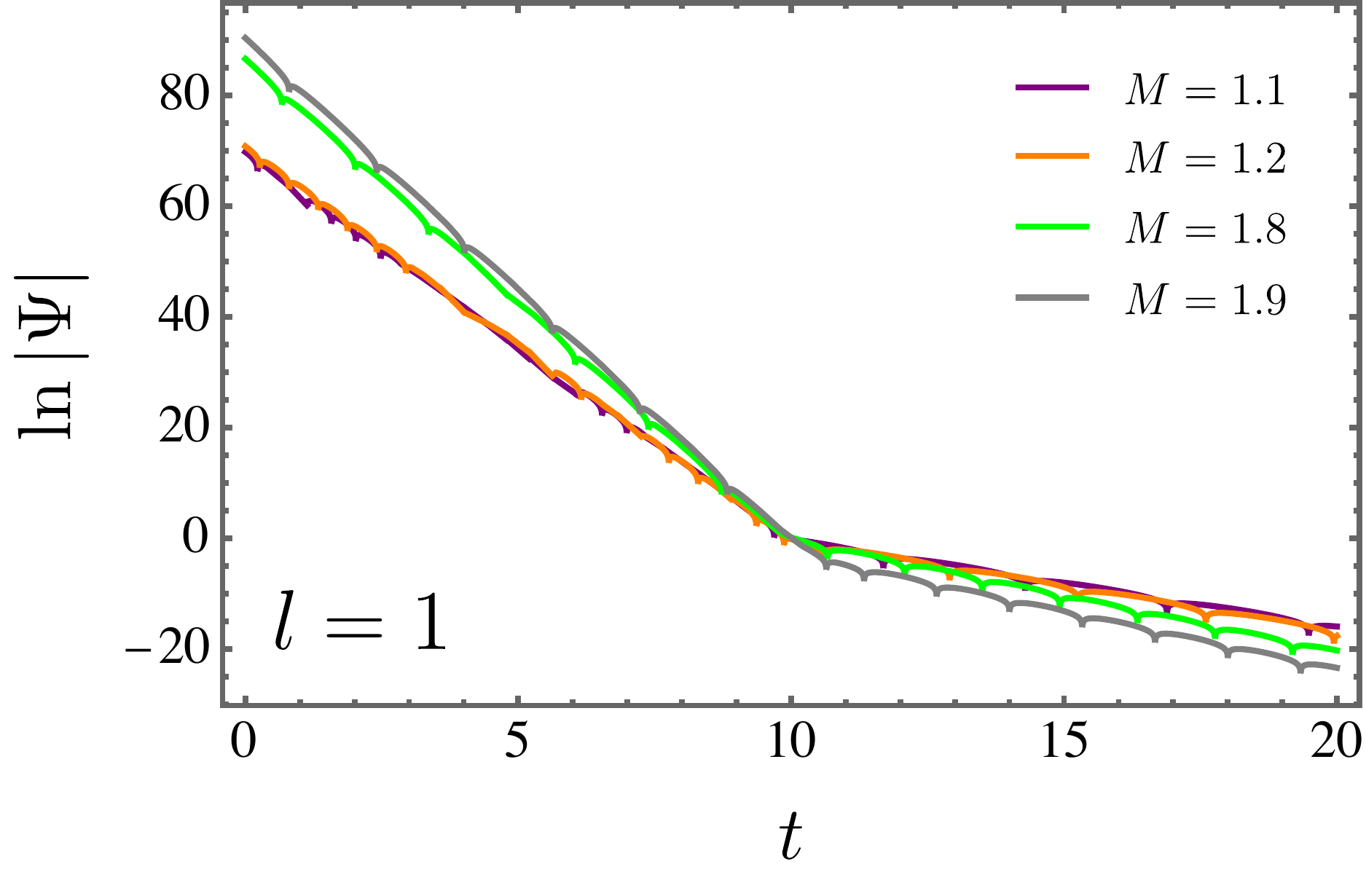}
    
    \caption{Time domain profiles of scalar perturbations at $r^{*} = 10M$ are shown for distinct values mass $M$ and $l$.}
    \label{Psiln}
\end{figure}

To corroborate our results, we provide Fig. \ref{Psiln}, which displays representative evolution profiles for different combinations of $M$ and $l$. 


\section{Conclusion}

This study delved into the examination of a regular black hole within the framework of Verlinde's emergent gravity, specifically focusing on the modified Hayward--like solution. In other words, we intended to see the signatures/direct consequences of the modifications of this spacetime ascribed to the dark matter effects.

The investigation revealed intriguing facets of this particular black hole configuration. Notably, under specific conditions, such as when the mass parameter $M$ exceeded a threshold of $0.13705$ (assuming $a_{0}=a=1$), a singular event horizon emerged. However, when we maintained $M$ at $1$ while exploring the range $0 < a \leqslant 0.31977$, a striking phenomenon occurred: three distinct horizons appeaed, denoted as $r_{+}$, $r_{\cdot}$, and $r_{-}$. The former was visually represented as the event horizon, whereas the latter were identified as the Cauchy horizons.

Our examination of the Hawking temperature and heat capacity unveiled the presence of at least two distinct phase transitions occurring around $M \approx 0.054$, and $M \approx 0.249$. Notably, in contrast to the conventional Schwarzschild black hole, this configuration exhibited a stable region within the parameter range of $M \approx 0.508-0.547$. This stability could potentially be attributed to the influence of dark matter within the system under investigation.

Furthermore, we explored three different approaches to compute the Hawking temperature: via surface gravity, through Hawking radiation, and by applying the first law of thermodynamics. Notably, it was pointed out in \cite{ma2014corrected,maluf2018thermodynamics} that the temperature derived from the last method did not yield a correct value for entropy calculations. To address this issue, we adopted the approach outlined in Ref. \cite{ma2014corrected}, which advocates for a correction factor encapsulated by $\Upsilon(r_{+},a)$ in our analysis.

Our computations of geodesic trajectories and critical orbits, specifically photon spheres, yielded significant insights into the spacetime configuration encompassing the black hole. Remarkably, we identified the existence of three distinct light rings denoted as $r_{c_{-}}$, $r_{c_{\asymp}}$, and $r_{c_{+}}$, signifying regions where photons can maintain stable orbits. Additionally, we also explored the examination of the shadows cast by these phenomena.

On the other hand, by exploring the \textit{quasinormal} modes through the utilization of third-- and sixth--order WKB approximations, stable and unstable oscillations were observed for specific frequencies. In addition, the study extended its scope to investigate the phenomenon of time--dependent scattering in this particular scenario.

In terms of future perspectives, an intriguing analysis that deserves consideration is the exploration of quantum tunneling radiation and its corresponding backreaction effects, similar to the work conducted by the authors in Ref. \cite{silva2013quantum}. These and other ideas are now under development.



\section*{Acknowledgments}
\hspace{0.5cm}

Most of the calculations were performed by using the \textit{Mathematica} software. A. A. Araújo Filho is supported by Conselho Nacional de Desenvolvimento Cient\'{\i}fico e Tecnol\'{o}gico (CNPq) and Fundação de Apoio à Pesquisa do Estado da Paraíba (FAPESQ) -- [200486/2022-5] and [150891/2023-7]. In addition, the author is indebit with R. V. Maluf, G. J. Olmo, P. J. Porfírio, A. Yu. Petrov, K. Jusufi, and Meng--Sen Ma for the fruitful discussions and clarifications in the revised version of the manuscript.

\section{Data Availability Statement}

Data Availability Statement: No Data associated in the manuscript


\bibliographystyle{ieeetr}
\bibliography{main}

\end{document}